\documentclass[superscriptaddress,onecolumn, reprint]{revtex4-2}
\usepackage{graphicx}
\usepackage{physics}
\usepackage{amssymb}
\usepackage{epstopdf}
\usepackage{amsmath,dsfont,commath}
\usepackage{bm}
\usepackage{caption}
\usepackage{changes}
\usepackage{enumitem} 
\usepackage{braket}
\usepackage{xr}
\usepackage{algorithm} 
\usepackage{algpseudocode}
\usepackage{lipsum, siunitx} 
\usepackage{hyperref}

\usepackage{xr}

\makeatletter
\newcommand*{\addFileDependency}[1]{
\typeout{(#1)}
\@addtofilelist{#1}
\IfFileExists{#1}{}{\typeout{No file #1.}}
}
\makeatother

\newcommand*{\myexternaldocument}[1]{%
\externaldocument{#1}%
\addFileDependency{#1.tex}%
\addFileDependency{#1.aux}%
}

\myexternaldocument{siaux} 

\usepackage[capitalise]{cleveref}
\newcommand{\be}{\begin{equation}}
\newcommand{\ee}{\end{equation}}
\newcommand{\bea}{\begin{align}} 
\newcommand{\eea}{\end{align}}

\newcommand{\ba}{\begin{array}}
\newcommand{\ea}{\end{array}}

\newcommand{\mc}{\mathcal}

\newcommand{\eq}[1]{Eq. \eqref{#1}}

\newcommand{\half}{\frac{1}{2}}

\newcommand{\fs}[1]{\SI{#1}{\femto\second}}

\begin{document}
\title{ Coarse-Grained Geometric Quantum Dynamics in the Tensor Network Representation}
\author{Mo Sha}
\author{Bing Gu}
\email{gubing@westlake.edu.cn}
\affiliation{Department of Chemistry and Department of Physics, Westlake University, Hangzhou, Zhejiang, China, 310030}

\begin{abstract}
Quantum geometrical molecular dynamics provides a quantum geometric picture for understanding reactive dynamics, especially excited-state conical intersection dynamics, and also a numerically exact method for strongly correlated electron-nuclear dynamics. However, there are substantial challenges in describing medium-size molecules with tens of nuclear degrees of freedom. The main challenge is that it uses a discrete variable representation to discretize the molecular configuration space, and thus requires a tremendous number of quantum chemistry calculations to construct the electronic overlap matrix. Moreover, the expansion coefficients scale exponentially with molecular size for direct-product basis sets. We address these challenges by first introducing a coarse-grained local diabatic ansatz, followed by a tensor network representation of the expansion coefficients and the molecular time-evolution operator. With a full 24-dimensional demonstration using the pyrazine molecule, we show that such developments provide a highly accurate and computationally tractable method for high-dimensional, fully quantum, strongly coupled electron-nuclear dynamics from first principles.

\end{abstract}

\maketitle

The conventional understanding of chemistry is based on the Born--Oppenheimer approximation for ground-state calculation and the Born--Huang framework for nonadiabatic excited-state processes, exploiting the timescale separation between electrons and nuclei. In the Born--Oppenheimer dynamics, the nuclear motion is entirely determined by the landscape of the adiabatic potential energy surface (APES). non--Born--Oppenheimer effects, including nonradiative electronic transitions, the geometric phase effect, and diagonal Born--Oppenheimer corrections essential for excited-state conical intersection dynamics, are accounted for by first- and second-order derivative couplings and vector potentials \cite{MEAD1992geometric}. However, all such corrections, despite being of order $\mc{O}(M^{-1})$, diverge at electronic degeneracies, particularly at conical intersections.

Conical intersections are hypersurfaces on the APESs of polyatomic molecules where two or more electronic states become degenerate. They can act as efficient ``molecular funnels'' mediating ultrafast non-radiative transitions on the femtosecond ($10^{-15}$ \si{\second}) timescale and play a fundamental role in photophysics, photochemistry, and photobiology, such as in the primary event of vision, the photostability of DNA molecules, and energy conversion in photosynthesis \cite{DOMCKE2011Conical, GARCIA-VIDAL2021Manipulating, LARSON2020Conical, polak2020, polli2010, rafiq2023, WORNER2011Conical}.
Moreover, the geometric phase effects can lead to destructive interference in the nuclear motion that can significantly influence the outcome of reactive scattering, cold chemistry, and photodissociation~\cite{RYABINKIN2017Geometric,WANG2024Impact,XIE2017Constructive,KENDRICK1997Geometric, XIE2016Nonadiabatic,YUAN2018Observation}. 
To model and predict these ultrafast conical intersection dynamics, it is essential to go beyond the Born--Oppenheimer approximation and takes fully account of the strong electron-nuclear correlation in the course of dynamics.

The singular derivative couplings make the Born-Huang framework inappropriate for CI dynamics. 
One approach for treating conical intersection dynamics is through an adiabatic-to-diabatic transformation. 
Strict diabatization is impossible for a finite number of electronic states due to geometric obstruction \cite{MEAD1982Conditions}. 
Instead, many quasi-diabatization schemes depending on different criteria, which cannot be uniquely defined and are often not straightforward to use, have to be employed to construct a diabatic model, followed by accurate quantum wavepacket propagation on this diabatic potential energy surface \cite{ZHOU2019QuasiDiabatic, YARKONY2019Diabatic, FEDOROV2019discontinuous, SUBOTNIK2008Constructing, CHOI2020Which, TANNOR2007Introduction, ZHU2015Construction}.
However, these methods face a series of severe challenges in practical applications. 
One of the most fundamental obstacles is the ``curse of dimensionality'': the computational resources (memory and computation time) required to accurately describe a molecular wavefunction grow exponentially with the number of degrees of freedom. 
This makes full-dimensional, numerically exact simulations infeasible for all but the smallest molecules (typically with fewer than six degrees of freedom). 
A simpler approach is to construct a vibronic coupling model Hamiltonian, usually based on the crude adiabatic representation. 
The advantage is that it is possible to treat such analytical models with tens of nuclear degrees of freedom using a tensor network representation of the nuclear wavefunction, as in the multi-configuration time-dependent Hartree and time-dependent density-matrix renormalization group \cite{WORTH2004Multidimensional, WANG2003Multilayer, CAZALILLA2002Timedependent, SCHOLLWOCK2011densitymatrix, SCHOLLWOCK2011densitymatrixa}.
While such methods are powerful for high-dimensional quantum dynamics, their application requires pre-constructed, global diabatic model Hamiltonians. 
Thus, this approach mainly focuses on photophysical processes involving only small-amplitude motion, such as in molecular aggregates and crystals \cite{ALEOTTI2021Parameterization,MANDAL2018QuasiDiabatic,GUAN2021Highfidelity}.

Quantum geometrical molecular dynamics is a quantum geometric framework for both ground-state and excited-state chemistry that emphasizes a global view of the molecular fiber bundle \cite{XIE2025Quantum}. It retains the chemically highly intuitive and useful concept of APES but differs fundamentally from the Born--Huang framework in how the non--Born--Oppenheimer effects are described. In this geometric framework, \emph{all} non--Born--Oppenheimer effects are unified into a single term---the global overlap matrix of adiabatic electronic states. Unlike the Born--Huang representation, which suffers from singular derivative couplings, the overlap matrix remains bounded throughout the entire configuration space, even in systems with conical intersections.
Therefore, the quantum geometrical molecular dynamics provides a numerically exact and singularity-free method for modeling non-adiabatic conical intersection dynamics. 
The geometric approach is based on the local diabatic ansatz, with the nuclear motion described by a discrete variable representation (DVR) and the electronic motion by adiabatic states at the DVR grid points\cite{GU2023DiscreteVariable, GU2024Nonadiabatic, SHA2025Exponential}.  Note that the local diabatic ansatz is not a quasi-diabatization scheme but directly uses the electronic eigenstates. One of the main practical advantages is that it does not even require a smooth gauge fixing of the electronic states \cite{ZHU2024Making}, making it straightforward for \textit{ab initio} simulations because the output of any quantum chemistry method inevitably carries random signs (or phases) depending on the gauge group\cite{MEAD1979determination, MEAD1992geometric}. 
Despite its conceptual and practical advantages for strongly correlated electron-nuclear dynamics, the computational cost scales exponentially with the number of nuclear degrees of freedom. Prior work has sought to mitigate this issue by employing Smolyak sparse-grid discretization to reduce the number of grid points and linked product approximation to reduce costly electronic-structure calculations \cite{XIE2025Linked, XIE2024NondirectProduct}. However, the geometric approach is still computationally infeasible for medium-size molecules with dozens of degrees of freedom.

Here we introduce a coarse-grained local diabatic ansatz that is suitable for high-dimensional nonadiabatic quantum molecular dynamics with only adiabatic electronic input. 
Furthermore, within the coarse-grained ansatz, we employ a tensor network representation—specifically, the matrix product state for the expansion coefficients and the matrix product operator for the total molecular time-evolution operator.
We first benchmark its accuracy and efficiency against numerically exact  quantum geometrical molecular dynamics calculations on a reduced four-mode model of internal conversion in pyrazine. The power of our approach is then demonstrated by applying it to the full 24-mode pyrazine model, showcasing its ability to tackle complex, high-dimensional conical intersection problems.

\emph{Quantum Geometrical Molecular Dynamics---} In the quantum geometrical molecular dynamics,  the equation of motion is given by   \cite{XIE2025Quantum},
\be
\begin{split}
 i \frac{\partial \bm \chi(\bm R, t)}{\partial t} &= \int \dd \bm R' T(\bm R, \bm R') \bm A(\bm R, \bm R') \bm \chi(\bm R', t) \\ 
& + \bm V(\bm R) \bm \chi(\bm R, t)
\end{split}
\ee 
where $T(\bm R, \bm R') = \langle \bm R|\hat{T}_\text{N}|\bm R' \rangle$ is the coordinate representation of the kinetic energy operator.
The global electronic overlap matrix encodes the  quantum geometry of the molecular fiber bundle. 
Specifically, it is the generating function of all non-Abelian quantum geometric measures.
The first order contains the Berry connection and first-order derivative couplings, and the second order contains the  non-Abelian quantum geometric tensor. It can be generated to non-Hermitian Hamiltonians \cite{XIE2025Quantum}.

The integro-differential equation can be discretized by the local diabatic ansatz, 
\begin{equation}
    \Psi(\bm{r}, \bm{R},t) = \sum_{\bm{n}, \alpha} C_{\bm{n}\alpha} (t)\phi_\alpha(\bm{r}; \bm{R}_{\bm{n}}) \chi_{\bm{n}}(\bm{R}),
    \label{eq:ansatz}
\end{equation}
where $\chi_{\bm{n}}(\bm R) = \chi_{n_1}(R_1) \cdots \chi_{n_d}(R_d)$ with $\bm{n}=(n_1, \dots, n_d)$ is the discrete variable representation (DVR) basis set defining a set of grid points (i.e., configurations) $\{\bm{R}_{\bm{n}}\}$ \cite{GU2023DiscreteVariable, GU2024Nonadiabatic}, $\phi_\alpha(\bm r; \bm R_{\bm{n}})$ denotes the associated electronic eigenstates (i.e., eigenstates of the electronic Hamiltonian at $\bm R_{\bm{n}}$ with energies $V_{\bm{n} \alpha}$), and $C_{\bm{n}\alpha}(t)$ are the expansion coefficients.
The vibronic basis set is a direct product of a nuclear  basis function $\ket{\chi_{\bm{n}}}$ and adiabatic electronic states  $\ket{\phi_\alpha(\bm{R_n})}$ at a fixed geometry, thus providing a discrete local trivialization of the molecular fiber bundle consisting of the truncated electronic Hilbert space parameterized by the configuration space. 
We use a shorthand notation $\ket{\bm{n}\alpha} \equiv \ket{\phi_\alpha(\bm{R_n})} \otimes \ket{\chi_{\bm{n}}}$.
The local diabatic ansatz leverages the locality and orthonormality of the DVR basis to simplify the construction of electronic Hamiltonian matrix elements. This is particularly important for \textit{ab initio} modeling with the molecular Coulomb Hamiltonian, where the construction is significantly more challenging than for a model Hamiltonian.
The  electronic Hamiltonian are  diagonal in the DVR
 set$\braket{ \bm{m}\beta | \hat{H}_{\text{BO}}(\bm r; \bm R) | \bm{n}\alpha } = V_{\bm n \alpha} \delta_{\bm{m}\bm{n}} \delta_{\beta\alpha}$.
  A unique feature of the geometric approach is that the nuclear kinetic energy operator, $\hat{T}_\text{N}$,  acting solely on the nuclear space, is dressed by the electronic overlap matrix 
  \be \braket{ \bm{m}\beta | \hat{T}_\text{N} | \bm{n}\alpha } = A_{\bm{m}\beta,\bm{n}\alpha} T_{\bm{m}\bm{n}}
  \ee 
 Here, $T_{\bm{m}\bm{n}} = \braket{ \bm{m} | \hat{T}_\text{N} | \bm{n} }$ is the kinetic energy matrix element in the nuclear DVR basis, which often has an analytical form, and $A_{\bm{m}\beta,\bm{n}\alpha} = \braket{\phi_\beta(\bm{R_m}) | \phi_\alpha(\bm{R_n})}_{\bm{r}}$ is the electronic overlap matrix.
Inserting the ansatz \eq{eq:ansatz} into the time-dependent Schr\"{o}dinger equation, $i\partial_t\ket{\Psi(t)} = \hat{H}\ket{\Psi(t)}$, and left multiplying $\bra{\bm{m}\beta}$ yields the equation of motion 
\begin{equation}
    i\dot{C}_{\bm{m}\beta}(t) = V_\beta(\bm{R_m}) C_{\bm{m}\beta}(t) + \sum_{\bm{n},\alpha} A_{\bm{m}\beta,\bm{n}\alpha} T_{\bm{m}\bm{n}} C_{\bm{n}\alpha}(t).
    \label{eq:}
\end{equation}

With a direct-product nuclear basis set, there is an apparent challenge that the expansion coefficients, $C_{\bm{n}\alpha}(t)$, a $(d+1)$ dimensional tensor with shape $N_0 \times N_1 \times \cdots \times N_d$, where $d$ is the number of nuclear degrees of freedom, $N_0$ is the number of electronic states, and $N_k$ is the number of basis functions for the $k$-th coordinate, scale exponentially with the molecular size.
For this, we will employ an efficient tensor network representation.
However, the most severe challenge of geometric quantum dynamics comes from the electronic overlap matrix.
Even when the expansion coefficient can be expressed by a Hartree product, it does not reduce the cost associated with the electronic overlap matrix because the molecular configuration space remains exponentially large.
It is simply not possible to compute the electronic eigenstates for each configuration and construct the overlap matrix.
A naive idea is to reduce the number of grid points.
However, a dense grid is required to describe the nuclear dynamics because, upon photoexcitation, the electronic energy quickly flows into the nuclear degrees of freedom and the nuclear wave packet spreads quickly in the configuration space even when it is initially localized.

\emph{Coarse-Graining of the Electronic Space ---}
We realize that the nuclear grid plays dual roles in geometric quantum dynamics.
On one hand, it is used to represent the time-dependent nuclear wave packet.
This usually requires a dense grid, as the nuclear kinetic energy can be large in the course of excited-state dynamics.
On the other hand, the nuclear grid provides a local trivialization to describe the electronic quantum geometry.
The fineness required for this purpose is determined by how the electronic states vary with nuclear geometry\cite{FRANKEL2011Geometry, XIE2025Quantum}.
The dynamical and geometrical grids are enforced to be equivalent in the local diabatic ansatz.
The insight is that it is not necessary to use a very dense grid to describe the electronic quantum geometry, meaning there is substantial redundancy in the original ansatz.
While the molecular configuration space is necessarily exponentially large, the size of the effective electronic Hilbert space is most likely not exponentially large.

For many chemical reactions (e.g., photodissociation or photoisomerization), only a few reaction coordinates undergo large amplitude motion involving bond breaking and formation, whereas the rest undergo small amplitude motion.
It is expected that the electronic states do not vary significantly along those modes.
It is therefore convenient to partition the full set of nuclear coordinates $\bm{R} = \{\bm{q}, \bm{Q}\}$   into a few primary reactive coordinates, $\bm{q}$, that undergo large amplitude motion, and the remaining non-reactive coordinates, $\bm{Q}$, with small amplitude motion.
We introduce a coarse-grained local diabatic ansatz
\begin{equation}
\ket{\Psi( t)} = \sum_{\alpha, \bm n, \bm{\nu}} C_{\alpha \bm n \bm \nu}(t)  { \ket{\phi_{{\alpha}}(\bm{q_n}, \bm{Q}_0)}  \otimes \ket{\chi_{\bm{n}}}  \otimes \ket{ \xi_{\bm{\nu}}} },
	\label{eq:cg_basis}
\end{equation}
where $\ket{\xi_{\bm{\nu}}}, \bm \nu = (\nu_1, \nu_2, \dots, \nu_{d_\text{NR}})$ refers to the basis functions for the non-reactive coordinates.
This ansatz differs from the original ansatz in that only a small subset of electronic states corresponding to the nuclear configurations $(\bm{q_n}, \bm Q_0)$ are employed rather than the complete electronic manifold corresponding to $(\bm{q}_{\bm n},\bm{Q}_{\bm \nu})$.
This significantly reduces the electronic structure calculations given that the number of non-reactive modes is much larger than the number of reactive coordinates, $d_{\text{NR}} \gg d_\text{R}$.
In the coarse-grained vibronic basis set, the nuclear kinetic energy matrix elements remain intact as the nuclear basis functions are not changed.
However, the matrix elements of the electronic Hamiltonian $\braket{\beta \bm m \bm \nu | \hat{H}_\text{BO}(\bm r; \bm q, \bm Q) | \alpha \bm n \bm \nu}$ are not diagonal anymore as a consequence of the coarse-graining.
To compute its matrix elements, for the small-amplitude modes, we expand the electronic Hamiltonian at a reference point $\bm Q_0$ (up to the second order in this work):
\begin{equation}\label{H_expansion}
    \begin{split}
        \hat{H}_\text{BO}(\bm{r};\bm{q}, \bm{Q}) & =
        \hat{H}_0(\bm{r};\bm{q}) + \hat{\bm{F}}(\bm{r};\bm{q}) \cdot \Delta \hat{\bm{Q}} \\
        & + \frac{1}{2} \Delta \hat{\bm{Q}}^\top \hat{\bm{G}}(\bm{r};\bm{q}) \Delta \hat{\bm{Q}}  
        \equiv \hat{H}_0 + \hat{H}_1,
    \end{split}
\end{equation}
where $\hat{H}_0 \equiv \hat{H}_0(\bm{r};\bm{q}) = \hat{H}_\text{BO}(\bm{r};\bm{q}, \bm{Q}_0)$, 
$\hat{F}_k(\bm{r};\bm{q}) = \partial_k \hat{H}_{\text{BO}}(\bm{r};\bm{q}, \bm{Q}) \big|_{\bm{Q}_0}$, 
$\partial_k \equiv \partial/\partial Q_k$, 
and $\hat{G}_{k l}(\bm{r};\bm{q}) = \partial_k \partial_l \hat{H}_{\text{BO}}(\bm{r};\bm{q}, \bm{Q}) \big|_{\bm{Q}_0}$ 
are operators acting on the composite electronic and reactive coordinate space, 
and $\Delta \hat{\bm{Q}} = \hat{\bm{Q}} - \bm{Q}_0$. 
The total molecular Hamiltonian is thus partitioned as $\hat{H} = \hat{T}_\text{N} + \hat{H}_0 + \hat{H}_1$.

\emph{Matrix Product States and Operators---}
To circumvent the exponential growth of expansion coefficients, we represent the high-dimensional tensor $C_{\alpha \bm n}(t)$ in the tensor-train (TT) format, which is equivalently referred to as a matrix product state (MPS) in many-body physics \cite{SCHOLLWOCK2011densitymatrix,SCHOLLWOCK2011densitymatrixa}:
\begin{equation}
C_{\alpha \bm n }(t) = \sum_{a_0, \dots, a_{d-1}} {M}^{[0] \alpha}_{a_0}(t) {M}^{[1] n_0}_{a_0,a_1}(t) \cdots {M}^{[d] n_{d-1}}_{a_{d-1}}(t).
\end{equation}
Here, the first tensor $M^{[0]}$ is associated with the electronic state index $\alpha$, while the subsequent tensors $M^{[k]}$ ($k=1, \dots, d$) carry the physical indices $n_{k-1}$ corresponding to the nuclear degrees of freedom. The virtual indices $a_k$ (with bond dimension $D_k$) connect adjacent sites, encoding the entanglement structure. The MPS form can be constructed from the full tensor with controlled accuracy using the TT-SVD algorithm \cite{OSELEDETS2011TensorTrain}, as detailed in Algorithm~\ref{alg:ttsvd}. In this representation, we explicitly fix the inherent gauge degree of freedom to the right-canonical form. This step is essential to avoid amplified perturbation errors and reduced accuracy associated with non-canonical forms \cite{ZHANG2020Stability, SCHOLLWOCK2011densitymatrix}. Beyond gauge fixing, the ordering of sites is crucial for efficiency; strongly coupled degrees of freedom should be placed in proximity \cite{CHAN2002Highly, KINJO2025Permutation, MURG2012Algebraic}. Guided by this principle, we order the sites as follows: the electronic states first, followed by the reactive coordinates $\bm{q}$, and finally the non-reactive coordinates $\bm{Q}$.

To propagate the wavefunction, we extend the tensor-train framework to operators by representing the $(2d+2)$-dimensional time-evolution operator $\hat{U}(\Delta t) = e^{-i\hat{H}\Delta t}$ as a Matrix Product Operator (MPO). As the operator-space counterpart to the MPS, the MPO expresses the propagator as a contracted chain of local tensors:
\begin{equation}
 U_{\beta \bm m, \alpha \bm n} = \sum_{a_0, \dots, a_d} {W}^{[0]\beta \alpha}_{ a_0} {W}^{[1]m_0 n_0}_{a_0 a_1} \cdots {W}^{[d]m_{d-1} n_{d-1}}_{a_{d-1}},
\end{equation}
where $\bm m$ and $\bm n$ include all degrees of freedom, both reactive and non-reactive in this equation. With a second-order Trotter-Suzuki decomposition \cite{SUZUKI1976Generalized}, the time-evolution operator is given by
\begin{equation}
 e^{-i\hat{H}\Delta t} = e^{-i\hat{T}_\text{N}\Delta t/2} e^{-i\hat{H}_0\Delta t/2} e^{-i\hat{H}_1\Delta t} e^{-i\hat{H}_0\Delta t/2} e^{-i\hat{T}_\text{N}\Delta t/2}.
\end{equation}

Each component in the propagator can be decomposed into the MPO form. 
Firstly, the propagator associated with the nuclear kinetic energy operator acting only on the nuclear degrees of freedom becomes
\begin{equation}
\begin{split}
\bm{U}_{T_\text{N}} & = \langle {\beta} \bm{m\mu}| e^{-i\hat{T}_\text{N}\Delta t} | {\alpha} \bm{n \nu} \rangle\\ 
&= A_{\bm{m} {\beta},\bm{n} {\alpha}} \langle \bm{m} | e^{-i\hat{T}_{\bm q}\Delta t} | \bm{n} \rangle 
\braket{ \bm \mu | e^{-i\hat{T}_{\bm Q}\Delta t} |\bm {\nu} }
\end{split}
\end{equation}
Similar to the original ansatz, the reactive coordinate kinetic energy operator is dressed by the electronic overlap matrix 
$A_{\bm{m}{\beta},\bm{n}{\alpha}} = \braket{\phi_{\beta}(\bm{q_m}, \bm{Q}_0) | \phi_{\alpha}(\bm{q_n}, \bm{Q}_0)}_{\bm{r}}$. 
The components $\langle \bm{m} | e^{-i\hat{T}_{\bm{q}}\Delta t} | \bm{n} \rangle$ and $\langle \bm{\mu} | e^{-i\hat{T}_{\bm{Q}}\Delta t} | \bm{\nu} \rangle$ are separable for each nuclear degree of freedom for rectilinear coordinates, yielding a $d$-site rank-one MPO $\left( \prod_{k=1}^{d_\text{R}}\langle m_k | e^{-i\hat{T}_k \Delta t} | n_k \rangle \right) \left( \prod_{l=1}^{d_\text{NR}}\langle \mu_l | e^{-i\hat{T}_l \Delta t} | \nu_l \rangle \right)$ 
The full MPO for $\bm{U}_{T_\text{N}}$ is then constructed by element-wise product of the MPO for the overlap matrix $\bm{A}$ with this rank-one MPO. The bond dimension of the final propagator MPO is $D_k = 1$ for $ k = d_\text{R}, \dots, d$  and is determined by the MPO representation of $\bm{A}$ for the first $d_\text{R}$ bonds.
A schematic representation of the MPO construction is provided in the Supporting Information (Fig.~\ref{fig:kinetic}).

Next, regarding the propagator for $\hat{H}_0$, the matrix representation of $\hat{H}_0(\bm{q})$ is diagonal in the coarse-grained ansatz by construction, as is the matrix representation of the corresponding propagator, $\bm{U}_{H_0}$, with elements $\langle \bm{m}\beta | e^{-i\hat{H}_0\Delta t} | \bm{n}\alpha \rangle = e^{-i V_{\bm{n}\alpha} \Delta t}\delta_{\bm{m}\bm{n}}\delta_{\beta\alpha}$, where the electronic energies $V_{\bm{n}\alpha}$ are obtained from quantum chemistry calculations. The resulting high-dimensional diagonal tensor, with elements $e^{-i V_{\bm{n}\alpha} \Delta t}$, acts as the identity on the nonreactive coordinates and is compressed into an MPO representation using TT-SVD.

Finally, the propagator $\hat{U}_{H_1} = e^{-i\hat{H}_{1}\Delta t}$ is constructed by splitting it into a product of exponentials for the individual terms.
\begin{equation}\label{eq:H1_prop}
e^{-i\hat{H}_{1}\Delta t} = \prod_{k} e^{-i\hat{F}_k \Delta \hat{Q}_{k} {\Delta t}}  \prod_{k,l} e^{-i\hat{G}_{kl} \Delta \hat{Q}_{k} \Delta \hat{Q}_{l} \Delta t/ 2} .
\end{equation}
The MPO form for the full propagator is constructed by multiplying the MPOs of each term in the product above, applying a TT-rounding step(\cref{alg:ttround}) after each contraction to control the bond dimension.
Each operator term in the exponent, $\hat{F}_k(\bm q) \Delta \hat{Q}_k$, is a tensor product of an operator acting on the reactive space and one acting on a single nonreactive nuclear coordinate. 
Its matrix elements in the nuclear basis are
$
    \langle \beta \bm{m} \bm{\mu} | \hat{F}_k \Delta \hat{Q}_k | \alpha \bm{n} \bm{\nu} \rangle 
    = F_{k, \bm{n}}^{\beta \alpha} \Delta Q_{k\bm{\mu}} \, \delta_{\bm{m}\bm{n}} \delta_{\bm{\mu}\bm{\nu}},
$
with
$
    F_{k, \bm{n}}^{\beta \alpha} = \int \phi_{\beta}^*(\bm{r}; \bm{q}_{\bm{n}}, \bm{Q}_0) \, \hat{F}_k(\bm{r}; \bm{q}_{\bm{n}}) \, \phi_{\alpha}(\bm{r}; \bm{q}_{\bm{n}}, \bm{Q}_0) \, \mathrm{d} \bm{r}.
$
Here, $\Delta Q_{k,\bm \mu} \equiv Q_{k,\bm \mu} - Q_{k,0}$ is the scalar displacement of the $k$-th degree of freedom.
Similarly, the matrix elements for the Hessian operator $\hat{\bm{G}}$ are
$ \langle \beta \bm{m} \bm{\mu} | \hat{G}_{kl} \Delta \hat{Q}_k \Delta \hat{Q}_l | \alpha \bm{n} \bm{\nu} \rangle 
    = G_{kl, \bm{n}}^{\beta \alpha} \Delta Q_{k,\mu} \Delta Q_{l,\mu} \, \delta_{\bm{m}\bm{n}} \delta_{\bm{\mu}\bm{\nu}},
$
with
$
    G_{kl, \bm{n}}^{\beta \alpha} = \int \phi_{\beta}^*(\bm{r}; \bm{q}_{\bm{n}}, \bm{Q}_0) \, \hat{G}_{kl}(\bm{r}; \bm{q}_{\bm{n}}) \, \phi_{\alpha}(\bm{r}; \bm{q}_{\bm{n}}, \bm{Q}_0) \, \mathrm{d} \bm{r}.
$
Note that only matrix elements between the same molecular configurations are required.
A schematic representation of the MPO construction is provided in the Supporting Information (Fig.~\ref{fig:H1}). 
To obtain the MPO for each individual propagator (e.g., $e^{-i{\Delta t}\,\hat{F}_k \Delta \hat{Q}_{k} }$), we first construct the MPO for the operator in the exponent (e.g., $-i{\Delta t}\,\hat{F}_k \Delta \hat{Q}_{k} $), followed by exponentiation using the scaling and squaring method combined with a Taylor expansion \cite{HIGHAM2005Scaling}.

\begin{figure}[htbp]
    \includegraphics[width=0.8\columnwidth]{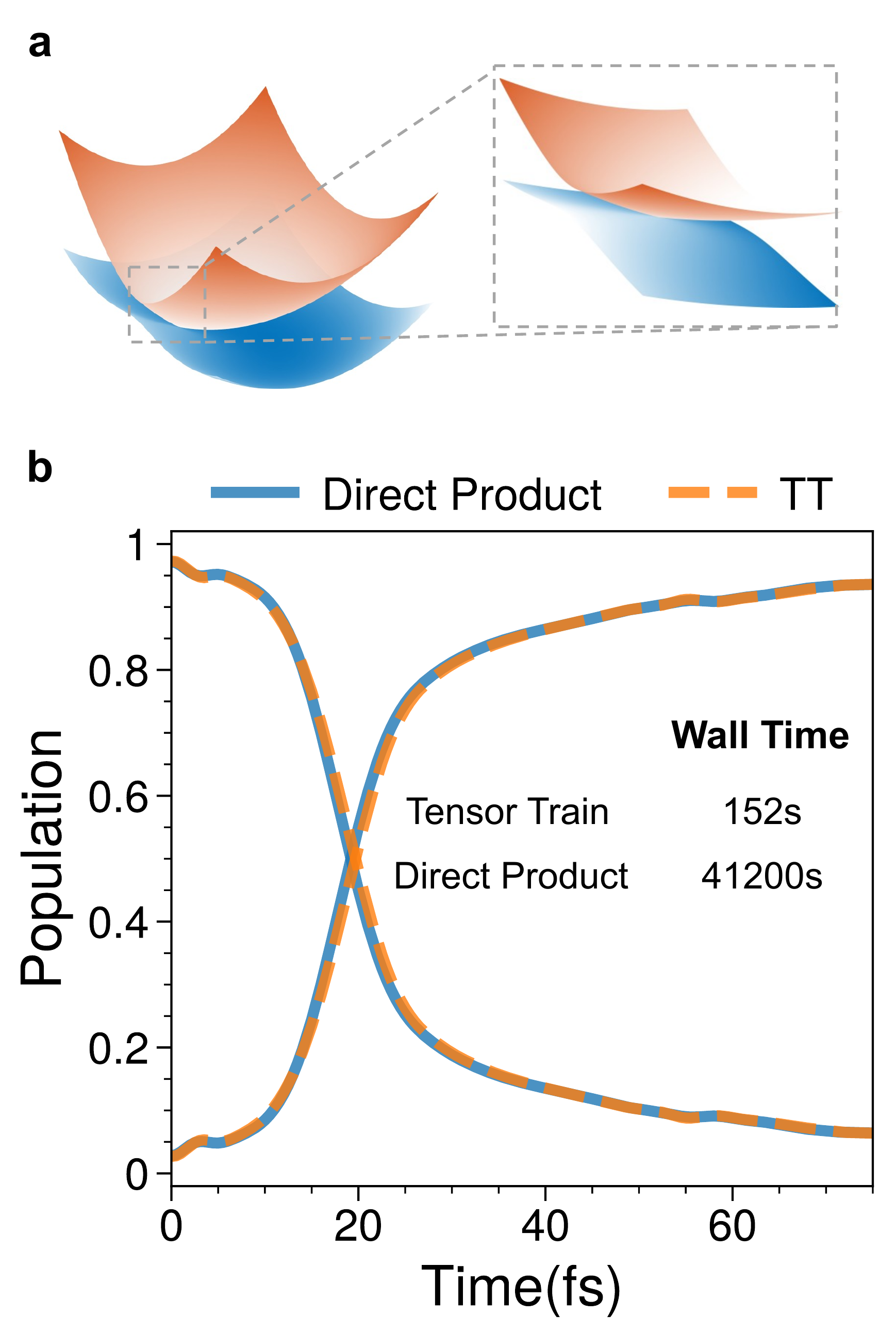}
    \caption{(a) The adiabatic potential energy surfaces of the $S_1$ and $S_2$ states of pyrazine. (b) Time evolution of the electronic population of two states. The result from the coarse-grained method (\mbox{$\epsilon=10^{-6}$}, \mbox{$D_{\max}=100$}, orange dashed line) is compared with the benchmark result from the direct product method (blue solid line) over the first \fs{75}.}
    \label{Fig:4mode}
\end{figure}
\begin{figure}[htbp]
    \includegraphics[width=0.9\columnwidth]{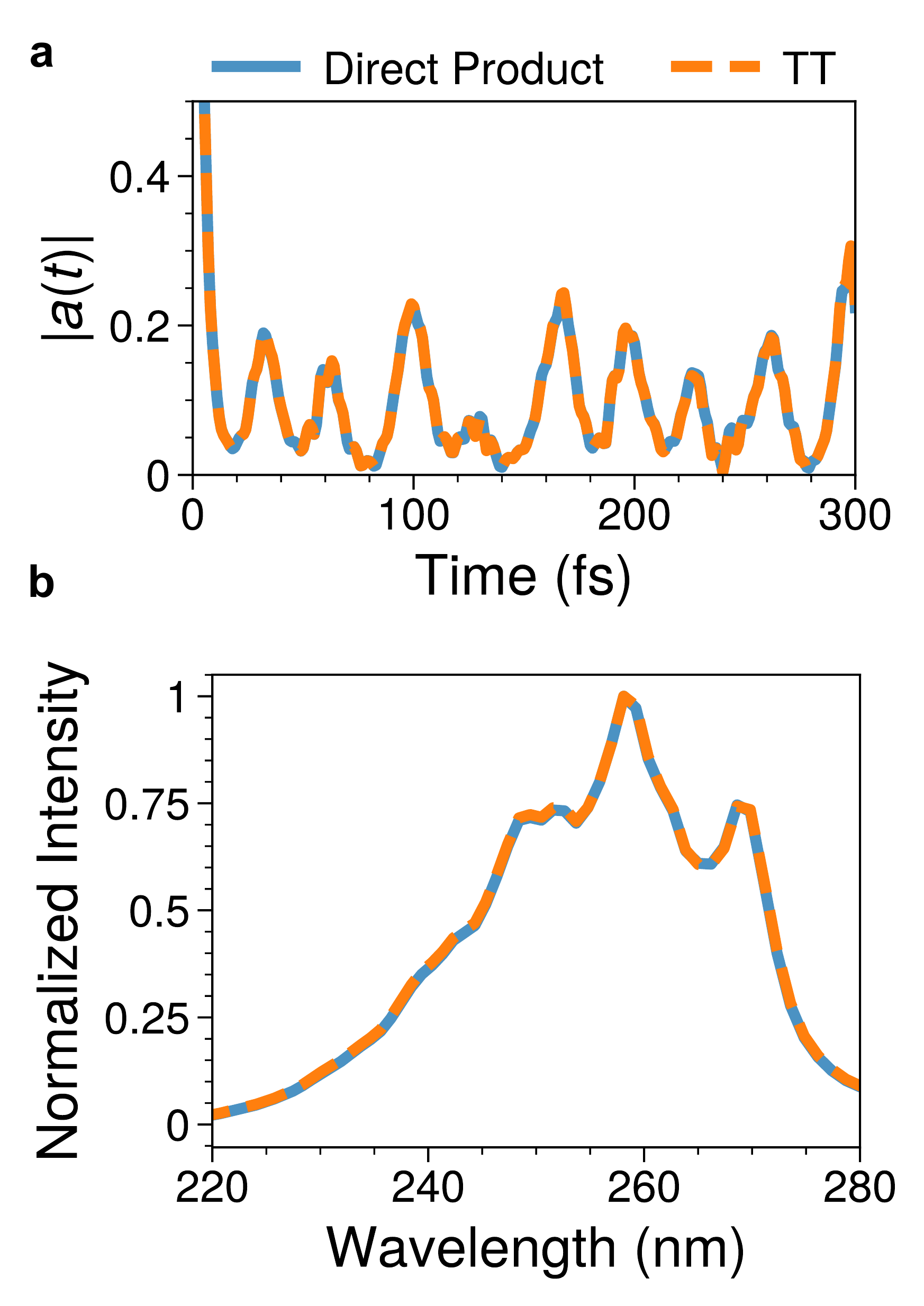}
    \caption{(a) The absolute value of the autocorrelation function, $|a(t)|$, obtained using the tensor train method compared with the benchmark result from the direct product method. (b) Comparison of the $S_2$ state absorption spectra calculated by the coarse-grained method and the direct product method.}
    \label{Fig:spectrum}
\end{figure}

We first demonstrate the accuracy and efficiency of this coarse-grained quantum geometrical molecular dynamics method in tensor-train representation, by simulating the internal conversion dynamics through a conical intersection in the pyrazine molecule following photoexcitation to the $S_2$ state.
It is straightforward to integrate any quantum chemistry method into our approach because a primary advantage of the quantum geometric framework is that the \emph{adiabatic} electronic eigenstates with random gauge choices from electronic structure calculations can be directly employed without gauge fixing or adiabatic-to-diabatic transformation.

Since current electronic structure software does not provide the required matrix elements of the derivative operators $\hat{\bm F}$ and $\hat{\bm G}$, we use a reduced two-state four-mode pyrazine model to create the adiabatic input (the adiabatic energies and the electronic overlap matrix). This model includes the most relevant three $A_g$ modes ($\nu_{6a}, \nu_{1}, \nu_{9a}$) and one $B_{1g}$ mode ($\nu_{10a}$) \cite{RAAB1999Molecular}.
For this reduced-scale simulation, a sine DVR basis\cite{COLBERT1992novel, LIGHT1985Generalized, LIGHT2000DiscreteVariablea} set with $N_k=15$ basis functions per mode and a time step of $\Delta t=\fs{0.5}$  were employed. In the coarse-grained ansatz, $\nu_{10a}$ and $\nu_{6a}$ were selected as the reactive modes. The rounding parameters for the MPS and MPO were set to an absolute error threshold of $\epsilon=10^{-6}$ and a maximum bond dimension of $D_{\max}=100$.

The simulation was run for \fs{150}, starting from a Gaussian nuclear wavepacket vertically excited to the $S_2$ state and centered at $\bm{R}=\bm{0}$, $\Psi_0=\prod_{i=1}^{d}({\pi^{-1/4}e^{-1R_i^2/2}})|S_2\rangle$, which is represented as a matrix product state . The reference geometry for the non-reactive modes was the Franck-Condon point ($\bm{Q}_0=\bm{0}$).
The coarse-grained method is in excellent agreement with the fine-grained direct product method for both the electronic population dynamics (\cref{Fig:4mode}) and the optical absorption spectrum (\cref{Fig:spectrum}). This accuracy is achieved with a significant computational speed up: the simulation with tensor network is two orders of magnitude faster than the direct product method (\SI{152}{\second} of wall time vs. \SI{41200}{\second}).
In \cref{Fig:spectrum}(a), the absolute value of the autocorrelation function is given by $a(t) = \langle \Psi_0 | \Psi(t) \rangle=\langle \Psi(t/2)^* | \Psi(t/2) \rangle$. The corresponding absorption spectrum presented in \cref{Fig:spectrum}(b) is calculated as the Fourier transform of the autocorrelation function: $I(\omega) = \int_{-\infty}^{\infty} \dd t\, a(t)e^{i\omega  t}e^{-|t|/\tau_h}\cos \left(\frac{\pi t}{2T}\right)$,where $\tau_h = \fs{30}$ is a damping time and $T$ is the total simulation time.

\begin{figure}[htbp]
    \includegraphics[width=0.8\columnwidth]{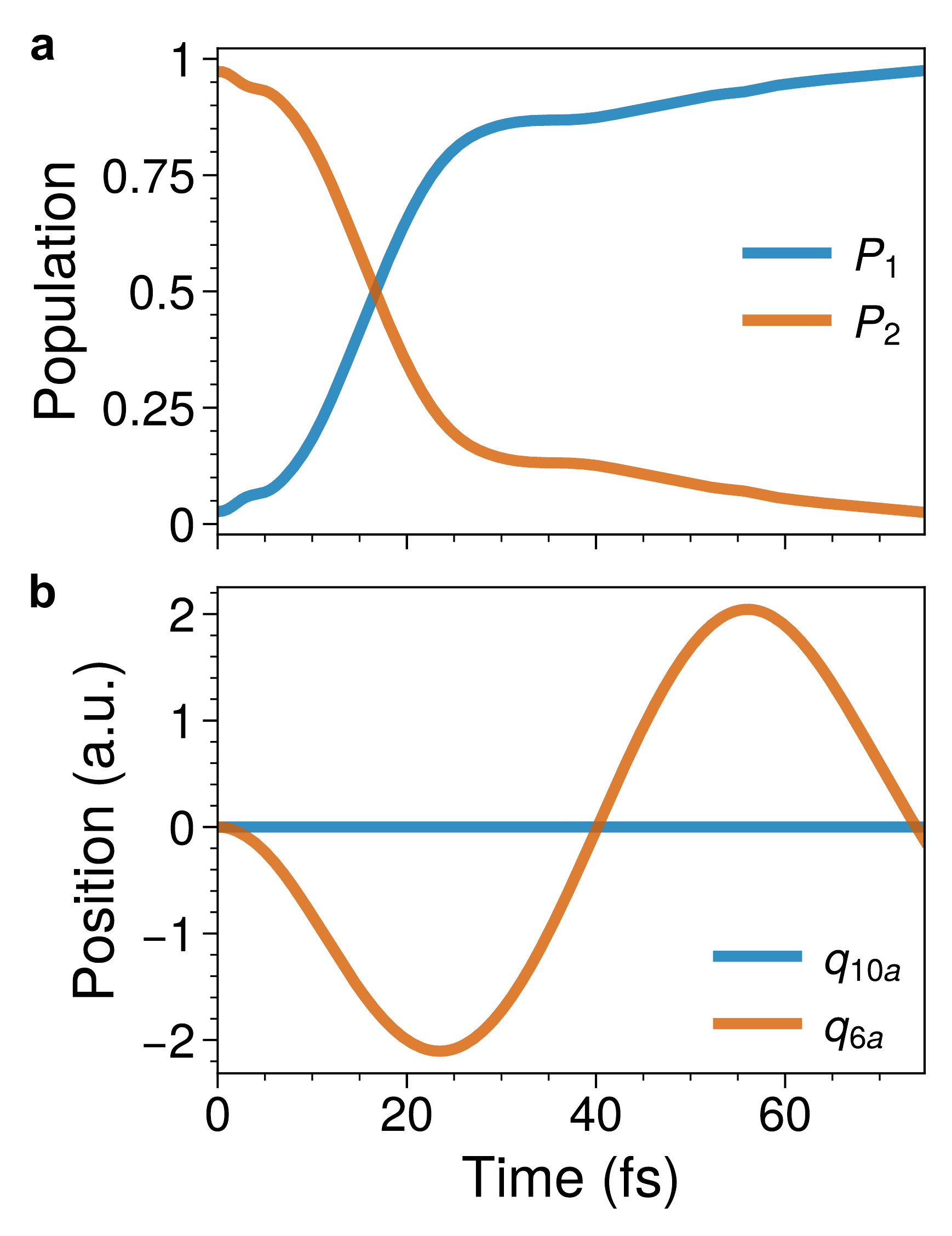}
    \caption{Nonadiabatic conical intersection dynamics  for the full 24-mode pyrazine model. {(a)} Electronic populations of the initially state $S_2$, and the lower-lying state $S_1$. {(b)} Expectation values of the position for the coupling mode $q_{10a}$, and the tuning mode $q_{6a}$. }
    \label{Fig:24mode}
\end{figure}
\begin{figure}[htbp]
    \includegraphics[width=1\columnwidth]{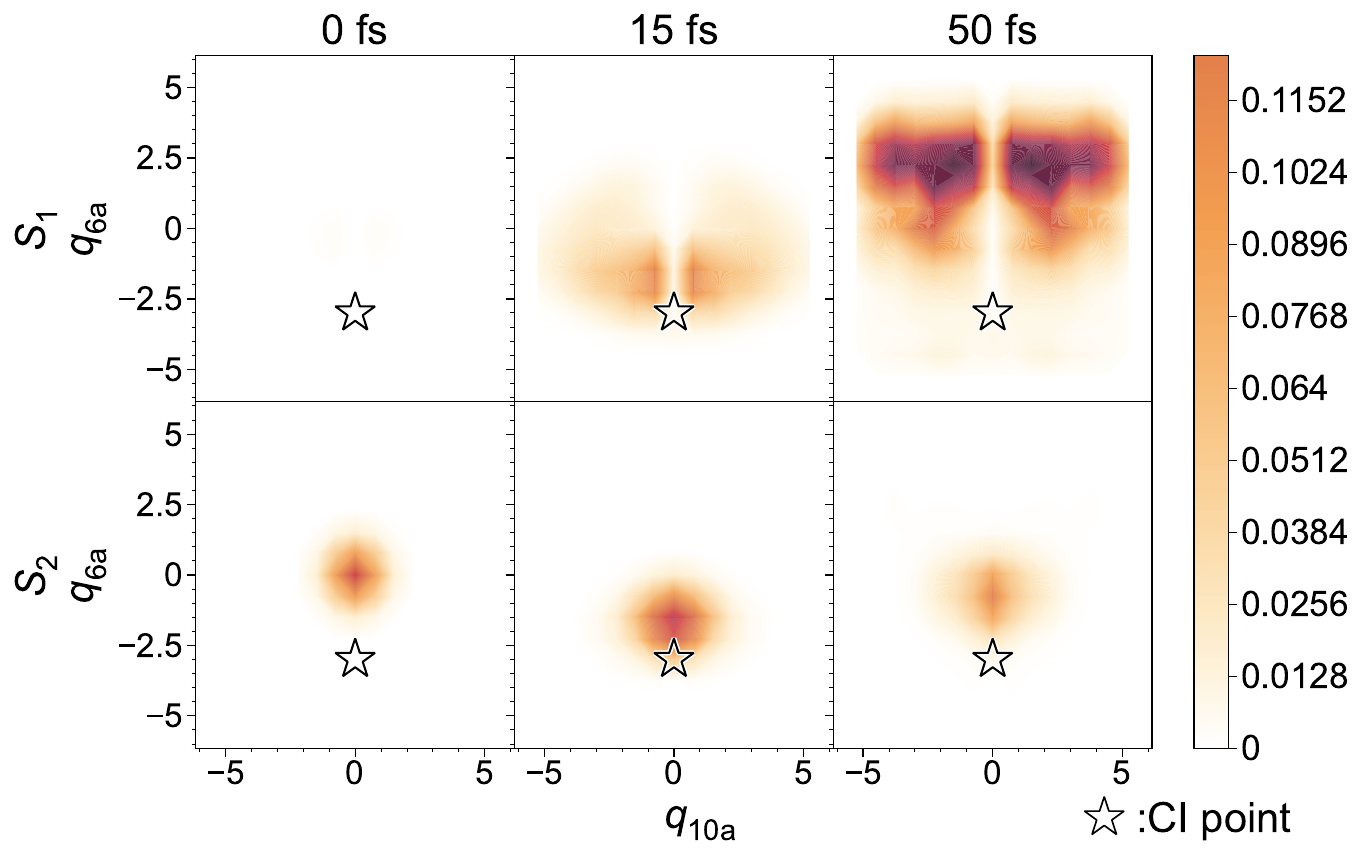}
    \caption{Reduced probability densities for the 24-mode pyrazine model on the $S_2$ (bottom) and $S_1$ (top) surfaces. The densities are projected onto the branching plane spanned by the coupling mode $q_{10a}$ and the tuning mode $q_{6a}$. The position of the conical intersection is indicated by a star. The wavepacket is initially excited to the $S_2$ state. At \fs{50}, the density transferred to the $S_1$ surface exhibits a nodal line along $q_{10a}=0$, a signature of the geometric phase effect induced by the conical intersection.}
    \label{Fig:24modepd}
\end{figure}

To demonstrate  the capability of our method for high-dimensional conical intersection dynamics, we now  apply it to the internal conversion dynamics in the full-dimensional pyrazine system with $d = 24$, far beyond the capability of direct-product method. 
The $\nu_{6a}$ and $\nu_{10a}$ modes are chosen as reactive modes and the rest as nonreactive modes. 
A sine DVR basis was used for all modes with $N_k = 15$ basis functions per mode. The time step was set to $\Delta t=\fs{0.5}$, the TT-rounding error threshold to $\epsilon=10^{-6}$, and the maximum bond dimension to $D_{\max} = 40$.
The reference geometry for the nonreactive modes was chosen as  $\bm{Q}_0=\bm{0}$ corresponding to the ground state minimum. 
The model Hamiltonian is only used to generate the adiabatic input including electronic energies along the reactive coordinates and the nuclear gradients $\bm F$ and $ \bm G$; its analytic form is not explicitly used in constructing the MPO.
We also start with a Gaussian  wavepacket centered at $\bm{R}=\bm{0}$ vertically excited to the $S_2$ state $\ket{\Psi_0} =(\prod_{i=1}^{d}{\pi^{-1/4}e^{-\half R_i^2}})|S_2\rangle$. 
As shown in \cref{Fig:24mode} (a), the coarse-grained method successfully captured the ultrafast electronic relaxation  dynamics from S$_2$ to S$_1$ within $\sim \fs{30}$ timescale.
The dynamics of the two reactive coordinates are shown in \cref{Fig:24mode} (b). The expectation value of  $\langle q_{6a}(t) \rangle$ (orange line), exhibits significant oscillatory motion, indicating that the nuclear wavepacket moves along this coordinate upon passing through the conical intersection. The wavepacket reaches its first maximum displacement at approximately \fs{25} and oscillates. 
In contrast, the expectation value of the coupling mode position, $\langle q_{10a}(t) \rangle$ (blue line), remains at 0 throughout the entire simulation due to the mirror symmetry of the model with respect to this coordinate, as expected. There is, nevertheless, significant diffusion along this mode occurring during the dynamics, see \cref{Fig:24modepd}.
For the reduced probability densities projected onto the two reactive coordinates $q_{6a}$ and $q_{10a}$,  $\rho_k(\bm{q}, t) = \int \left| \Psi_k(\bm{q},\bm{Q}, t) \right|^2 d\bm{Q}$. 
At $t=\fs{0}$, the wavepacket is centrally localized on the $S_2$ surface. As the population transfers to the $S_1$ surface, the wavepacket undergoes significant spreading. Crucially, at $t=\fs{50}$, the density formed on the $S_1$ surface exhibits a distinct double-lobe structure with a nodal plane along $q_{10a}=0$, where the probability density vanishes. This nodal feature is a manifestation of the geometric phase effects. 
Our method remains stable at longer times beyond the first passage through the conical intersection, see \cref{Fig:full24modelonger} for a longer time simulation up to \fs{150}, with 14 hours of wall time.

To summarize, we have introduced a coarse-grained geometric quantum dynamics framework in the tensor-network representation for high-dimensional strongly coupled electron-nuclear dynamics. 
It allows direct \textit{ab initio} simulation of conical intersection dynamics incorporating all non--Born--Oppenheimer effects including not only nonadiabatic coupling but also geometric phase effect, diagonal Born--Oppenheimer correction, and second-order derivative couplings.  It does not require a pre-constructed global diabatic analytic Hamiltonian or approximate quasi-diabatization to remove singular derivative couplings. 
The accuracy and efficiency of the method are demonstrated first by a four-dimensional pyrazine model followed by a full-dimensional nonadiabatic quantum molecular dynamics modeling. The results imply that the geometric quantum dynamics provides a powerful tool for direct modeling of high-dimensional quantum dynamics at conical intersections from first principles without constructing a diabatic model Hamiltonian. The coarse-graining scheme employed here is a simple version when reactive coordinates can be clearly identified; for more complex reactions, an automatic algorithm can be developed for the coarse-graining step.

\appendix
\bibliography{all,dynamics}

@article{ALEOTTI2021Parameterization,
  title = {Parameterization of a {{Linear Vibronic Coupling Model}} with {{Multiconfigurational Electronic Structure Methods}} to {{Study}} the {{Quantum Dynamics}} of {{Photoexcited Pyrene}}},
  author = {Aleotti, Flavia and Aranda, Daniel and Yaghoubi Jouybari, Martha and Garavelli, Marco and Nenov, Artur and Santoro, Fabrizio},
  year = 2021,
  month = mar,
  journal = {J. Chem. Phys.},
  volume = {154},
  number = {10},
  pages = {104106},
  publisher = {American Institute of Physics},
  issn = {0021-9606},
  doi = {10.1063/5.0044693},
  urldate = {2021-03-15},
  copyright = {All rights reserved}
}

@article{CAZALILLA2002Timedependent,
  title = {Time-Dependent Density-Matrix Renormalization Group: A Systematic Method for the Study of Quantum Many-Body out-of-Equilibrium Systems},
  shorttitle = {Time-Dependent Density-Matrix Renormalization Group},
  author = {Cazalilla, M. A.},
  year = 2002,
  journal = {Physical Review Letters},
  volume = {88},
  number = {25},
  doi = {10.1103/PhysRevLett.88.256403},
  langid = {english}
}

@article{CHAN2002Highly,
  title = {Highly Correlated Calculations with a Polynomial Cost Algorithm: {{A}} Study of the Density Matrix Renormalization Group},
  shorttitle = {Highly Correlated Calculations with a Polynomial Cost Algorithm},
  author = {Chan, Garnet Kin-Lic and {Head-Gordon}, Martin},
  year = 2002,
  month = mar,
  journal = {Journal of Chemical Physics},
  volume = {116},
  number = {11},
  pages = {4462--4476},
  issn = {0021-9606},
  doi = {10.1063/1.1449459},
  urldate = {2024-09-03}
}

@article{CHOI2020Which,
  title = {Which Form of the Molecular Hamiltonian Is the Most Suitable for Simulating the Nonadiabatic Quantum Dynamics at a Conical Intersection?},
  author = {Choi, Seonghoon and Van{\'i}{\v c}ek, Ji{\v r}{\'i}},
  year = 2020,
  month = dec,
  journal = {Journal of Chemical Physics},
  volume = {153},
  number = {21},
  pages = {211101},
  publisher = {American Institute of Physics},
  issn = {0021-9606},
  doi = {10.1063/5.0033410},
  urldate = {2025-09-28},
  langid = {english}
}

@article{COLBERT1992novel,
  title = {A Novel Discrete Variable Representation for Quantum Mechanical Reactive Scattering via the {{{\emph{S}}}} -Matrix {{Kohn}} Method},
  author = {Colbert, Daniel T. and Miller, William H.},
  year = 1992,
  month = feb,
  journal = {Journal of Chemical Physics},
  volume = {96},
  number = {3},
  pages = {1982--1991},
  issn = {0021-9606, 1089-7690},
  doi = {10.1063/1.462100},
  urldate = {2025-06-16},
  langid = {english}
}

@book{DOMCKE2011Conical,
  title = {Conical {{Intersections}}: {{Theory}}, {{Computation}} and {{Experiment}}},
  shorttitle = {Conical {{Intersections}}},
  author = {Domcke, Wolfgang and Yarkony, David R. and K{\"o}ppel, Horst},
  year = 2011,
  publisher = {World Scientific},
  googlebooks = {Fc1bN093a4wC},
  isbn = {978-981-4313-45-2},
  langid = {english}
}

@article{ECKART1936Approximation,
  title = {The {{Approximation}} of {{One Matrix}} by {{Another}} of {{Lower Rank}}},
  author = {Eckart, Carl and Young, Gale},
  year = 1936,
  month = sep,
  journal = {Psychometrika},
  volume = {1},
  number = {3},
  pages = {211--218},
  issn = {0033-3123, 1860-0980},
  doi = {10.1007/BF02288367},
  urldate = {2025-09-10},
  langid = {english}
}

@article{FEDOROV2019discontinuous,
  title = {A Discontinuous Basis Enables Numerically Exact Solution of the {{Schr\"odinger}} Equation around Conical Intersections in the Adiabatic Representation},
  author = {Fedorov, Dmitry A. and Levine, Benjamin G.},
  year = 2019,
  month = feb,
  journal = {Journal of Chemical Physics},
  volume = {150},
  number = {5},
  pages = {054102},
  issn = {0021-9606},
  doi = {10.1063/1.5058268},
  urldate = {2023-07-22}
}

@book{FRANKEL2011Geometry,
  title = {The {{Geometry}} of {{Physics}}: {{An Introduction}}},
  shorttitle = {The {{Geometry}} of {{Physics}}},
  author = {Frankel, Theodore},
  year = 2011,
  edition = {3},
  publisher = {Cambridge University Press},
  address = {Cambridge},
  doi = {10.1017/CBO9781139061377},
  urldate = {2026-01-13},
  isbn = {978-1-107-60260-1}
}

@article{GARCIA-VIDAL2021Manipulating,
  title = {Manipulating {{Matter}} by {{Strong Coupling}} to {{Vacuum Fields}}},
  author = {{Garcia-Vidal}, Francisco J. and Ciuti, Cristiano and Ebbesen, Thomas W.},
  year = 2021,
  month = jul,
  journal = {Science},
  volume = {373},
  number = {6551},
  pages = {eabd0336},
  publisher = {American Association for the Advancement of Science},
  issn = {0036-8075, 1095-9203},
  doi = {10.1126/science.abd0336},
  urldate = {2021-07-09},
  chapter = {Review},
  copyright = {Copyright \textbackslash copyright 2021 The Authors, some rights reserved; exclusive licensee American Association for the Advancement of Science. No claim to original U.S. Government Works. https://www.sciencemag.org/about/science-licenses-journal-article-reuseThis is an article distributed under the terms of the Science Journals Default License.},
  langid = {english}
}

@article{GU2023DiscreteVariable,
  title = {A {{Discrete-Variable Local Diabatic Representation}} of {{Conical Intersection Dynamics}}},
  author = {Gu, Bing},
  year = 2023,
  month = oct,
  journal = {Journal of Chemical Theory and Computation},
  volume = {19},
  number = {19},
  pages = {6557--6563},
  publisher = {American Chemical Society},
  issn = {1549-9618, 1549-9626},
  doi = {10.1021/acs.jctc.3c00560},
  urldate = {2025-06-12},
  copyright = {https://doi.org/10.15223/policy-029},
  langid = {english}
}

@article{GU2024Nonadiabatic,
  title = {Nonadiabatic {{Conical Intersection Dynamics}} in the {{Local Diabatic Representation}} with {{Strang Splitting}} and {{Fourier Basis}}},
  author = {Gu, Bing},
  year = 2024,
  month = apr,
  journal = {Journal of Chemical Theory and Computation},
  volume = {20},
  number = {7},
  pages = {2711--2718},
  publisher = {American Chemical Society},
  issn = {1549-9618, 1549-9626},
  doi = {10.1021/acs.jctc.3c01317},
  urldate = {2025-06-12},
  copyright = {https://doi.org/10.15223/policy-029},
  langid = {english}
}

@article{GUAN2021Highfidelity,
  title = {High-Fidelity First Principles Nonadiabaticity: Diabatization, Analytic Representation of Global Diabatic Potential Energy Matrices, and Quantum Dynamics},
  shorttitle = {High-Fidelity First Principles Nonadiabaticity},
  author = {Guan, Yafu and Xie, Changjian and Yarkony, David R. and Guo, Hua},
  year = 2021,
  month = nov,
  journal = {Physical Chemistry Chemical Physics},
  volume = {23},
  number = {44},
  pages = {24962--24983},
  publisher = {The Royal Society of Chemistry},
  issn = {1463-9084},
  doi = {10.1039/D1CP03008F},
  urldate = {2026-01-08},
  langid = {english}
}

@article{HIGHAM2005Scaling,
  title = {The {{Scaling}} and {{Squaring Method}} for the {{Matrix Exponential Revisited}}},
  author = {Higham, Nicholas J.},
  year = 2005,
  month = jan,
  journal = {SIAM Journal on Matrix Analysis and Applications},
  volume = {26},
  number = {4},
  pages = {1179--1193},
  issn = {0895-4798, 1095-7162},
  doi = {10.1137/04061101X},
  urldate = {2025-09-12},
  langid = {english}
}

@article{KENDRICK1997Geometric,
  title = {Geometric {{Phase Effects}} in the {{Vibrational Spectrum}} of {{Na}} 3 ( {{X}} )},
  author = {Kendrick, Brian},
  year = 1997,
  month = sep,
  journal = {Physical Review Letters},
  volume = {79},
  number = {13},
  pages = {2431--2434},
  issn = {0031-9007, 1079-7114},
  doi = {10.1103/PhysRevLett.79.2431},
  urldate = {2025-12-24},
  copyright = {http://link.aps.org/licenses/aps-default-license},
  langid = {english}
}

@article{KINJO2025Permutation,
  title = {Permutation of {{Tensor-Train Cores}} for {{Computing Moments}} on {{Stochastic Differential Equations}}},
  author = {Kinjo, Kayo and Sakurai, Rihito and Kishimoto, Tatsuya and Ohkubo, Jun},
  year = 2025,
  month = aug,
  journal = {Journal of the Physical Society of Japan},
  volume = {94},
  number = {8},
  eprint = {2504.10492},
  primaryclass = {physics},
  pages = {084001},
  issn = {0031-9015, 1347-4073},
  doi = {10.7566/JPSJ.94.084001},
  urldate = {2026-01-08},
  archiveprefix = {arXiv}
}

@book{LARSON2020Conical,
  title = {Conical {{Intersections}} in {{Physics}}: {{An Introduction}} to {{Synthetic Gauge Theories}}},
  shorttitle = {Conical {{Intersections}} in {{Physics}}},
  author = {Larson, Jonas and Sj{\"o}qvist, Erik and {\"O}hberg, Patrik},
  year = 2020,
  series = {Lecture {{Notes}} in {{Physics}}},
  volume = {965},
  publisher = {Springer International Publishing},
  address = {Cham},
  doi = {10.1007/978-3-030-34882-3},
  urldate = {2024-05-31},
  copyright = {http://www.springer.com/tdm},
  isbn = {978-3-030-34881-6 978-3-030-34882-3},
  langid = {english}
}

@article{LIGHT1985Generalized,
  title = {Generalized Discrete Variable Approximation in Quantum Mechanics},
  author = {Light, J. C. and Hamilton, I. P. and Lill, J. V.},
  year = 1985,
  month = feb,
  journal = {Journal of Chemical Physics},
  volume = {82},
  number = {3},
  pages = {1400--1409},
  issn = {0021-9606, 1089-7690},
  doi = {10.1063/1.448462},
  urldate = {2025-06-16},
  langid = {english}
}

@incollection{LIGHT2000DiscreteVariablea,
  title = {Discrete-{{Variable Representations}} and Their {{Utilization}}},
  booktitle = {Advances in {{Chemical Physics}}},
  author = {Light, John C. and Carrington, Tucker},
  editor = {Prigogine, I. and Rice, Stuart A.},
  year = 2000,
  month = jan,
  edition = {1},
  volume = {114},
  pages = {263--310},
  publisher = {Wiley},
  doi = {10.1002/9780470141731.ch4},
  urldate = {2025-06-16},
  isbn = {978-0-471-39267-5 978-0-470-14173-1},
  langid = {english}
}

@article{MANDAL2018QuasiDiabatic,
  title = {Quasi-{{Diabatic Representation}} for {{Nonadiabatic Dynamics Propagation}}},
  author = {Mandal, Arkajit and Yamijala, Sharma SRKC and Huo, Pengfei},
  year = 2018,
  month = apr,
  journal = {Journal of Chemical Theory and Computation},
  volume = {14},
  number = {4},
  pages = {1828--1840},
  publisher = {American Chemical Society},
  issn = {1549-9618},
  doi = {10.1021/acs.jctc.7b01178},
  urldate = {2023-04-05}
}

@article{MEAD1979determination,
  title = {On the Determination of {{Born}}--{{Oppenheimer}} Nuclear Motion Wave Functions Including Complications Due to Conical Intersections and Identical Nuclei},
  author = {Mead, C. Alden and Truhlar, Donald G.},
  year = 1979,
  month = mar,
  journal = {Journal of Chemical Physics},
  volume = {70},
  number = {5},
  pages = {2284--2296},
  issn = {0021-9606, 1089-7690},
  doi = {10.1063/1.437734},
  urldate = {2023-11-19},
  langid = {english}
}

@article{MEAD1982Conditions,
  title = {Conditions for the Definition of a Strictly Diabatic Electronic Basis for Molecular Systems},
  author = {Mead, C. Alden and Truhlar, Donald G.},
  year = 1982,
  month = dec,
  journal = {Journal of Chemical Physics},
  volume = {77},
  number = {12},
  pages = {6090--6098},
  issn = {0021-9606, 1089-7690},
  doi = {10.1063/1.443853},
  urldate = {2021-11-01},
  langid = {english}
}

@article{MEAD1992geometric,
  title = {The Geometric Phase in Molecular Systems},
  author = {Mead, C. Alden},
  year = 1992,
  month = jan,
  journal = {Reviews of Modern Physics},
  volume = {64},
  number = {1},
  pages = {51--85},
  publisher = {American Physical Society},
  doi = {10.1103/RevModPhys.64.51},
  urldate = {2025-08-20}
}

@article{MIRSKY1960SYMMETRIC,
  title = {{{SYMMETRIC GAUGE FUNCTIONS AND UNITARILY INVARIANT NORMS}}},
  author = {MIRSKY, L.},
  year = 1960,
  month = jan,
  journal = {The Quarterly Journal of Mathematics},
  volume = {11},
  number = {1},
  pages = {50--59},
  issn = {0033-5606},
  doi = {10.1093/qmath/11.1.50},
  urldate = {2025-09-10}
}

@article{MURG2012Algebraic,
  title = {Algebraic {{Bethe}} Ansatz and Tensor Networks},
  author = {Murg, V. and Korepin, V. E. and Verstraete, F.},
  year = 2012,
  month = jul,
  journal = {Physical Review B},
  volume = {86},
  number = {4},
  pages = {045125},
  publisher = {American Physical Society},
  doi = {10.1103/PhysRevB.86.045125},
  urldate = {2026-01-08}
}

@article{OSELEDETS2011TensorTrain,
  title = {Tensor-{{Train Decomposition}}},
  author = {Oseledets, I. V.},
  year = 2011,
  month = jan,
  journal = {SIAM Journal on Scientific Computing},
  volume = {33},
  number = {5},
  pages = {2295--2317},
  issn = {1064-8275, 1095-7197},
  doi = {10.1137/090752286},
  urldate = {2025-06-12},
  langid = {english}
}

@article{polak2020,
  title = {Manipulating Molecules with Strong Coupling: {{Harvesting}} Triplet Excitons in Organic Exciton Microcavities},
  shorttitle = {Manipulating Molecules with Strong Coupling},
  author = {Polak, Daniel and Jayaprakash, Rahul and Lyons, Thomas P. and {Mart{\'i}nez-Mart{\'i}nez}, Luis {\'A} and Leventis, Anastasia and Fallon, Kealan J. and Coulthard, Harriet and Bossanyi, David G. and Georgiou, Kyriacos and Anthony J. Petty, I. I. and Anthony, John and Bronstein, Hugo and {Yuen-Zhou}, Joel and Tartakovskii, Alexander I. and Clark, Jenny and Musser, Andrew J.},
  year = 2020,
  month = jan,
  journal = {Chemical Science},
  volume = {11},
  number = {2},
  pages = {343--354},
  publisher = {The Royal Society of Chemistry},
  issn = {2041-6539},
  doi = {10.1039/C9SC04950A},
  urldate = {2020-03-31},
  langid = {english}
}

@article{polli2010,
  title = {Conical Intersection Dynamics of the Primary Photoisomerization Event in Vision},
  author = {Polli, Dario and Alto{\`e}, Piero and Weingart, Oliver and Spillane, Katelyn M. and Manzoni, Cristian and Brida, Daniele and Tomasello, Gaia and Orlandi, Giorgio and Kukura, Philipp and Mathies, Richard A. and Garavelli, Marco and Cerullo, Giulio},
  year = 2010,
  month = sep,
  journal = {Nature},
  volume = {467},
  number = {7314},
  pages = {440--443},
  publisher = {Nature Publishing Group},
  issn = {1476-4687},
  doi = {10.1038/nature09346},
  urldate = {2022-01-17},
  copyright = {2010 Nature Publishing Group, a division of Macmillan Publishers Limited. All Rights Reserved.},
  langid = {english}
}

@article{RAAB1999Molecular,
  title = {Molecular Dynamics of Pyrazine after Excitation to the {{S2}} Electronic State Using a Realistic 24-Mode Model {{Hamiltonian}}},
  author = {Raab, A. and Worth, G. A. and Meyer, H.-D. and Cederbaum, L. S.},
  year = 1999,
  month = jan,
  journal = {Journal of Chemical Physics},
  volume = {110},
  number = {2},
  pages = {936--946},
  issn = {0021-9606, 1089-7690},
  doi = {10.1063/1.478061},
  urldate = {2025-09-15},
  langid = {english}
}

@article{rafiq2023,
  title = {Spin--Vibronic Coherence Drives Singlet--Triplet Conversion},
  author = {Rafiq, Shahnawaz and Weingartz, Nicholas P. and Kromer, Sarah and Castellano, Felix N. and Chen, Lin X.},
  year = 2023,
  month = aug,
  journal = {Nature},
  volume = {620},
  number = {7975},
  pages = {776--781},
  publisher = {Nature Publishing Group},
  issn = {1476-4687},
  doi = {10.1038/s41586-023-06233-y},
  urldate = {2023-11-20},
  copyright = {2023 The Author(s), under exclusive licence to Springer Nature Limited},
  langid = {english}
}

@article{RYABINKIN2017Geometric,
  title = {Geometric {{Phase Effects}} in {{Nonadiabatic Dynamics}} near {{Conical Intersections}}},
  author = {Ryabinkin, Ilya G. and {Joubert-Doriol}, Lo{\"i}c and Izmaylov, Artur F.},
  year = 2017,
  month = jul,
  journal = {Accounts of Chemical Research},
  volume = {50},
  number = {7},
  pages = {1785--1793},
  publisher = {American Chemical Society},
  issn = {0001-4842},
  doi = {10.1021/acs.accounts.7b00220},
  urldate = {2025-08-21},
  langid = {english}
}

@article{SCHOLLWOCK2011densitymatrix,
  title = {The Density-Matrix Renormalization Group: A Short Introduction},
  shorttitle = {The Density-Matrix Renormalization Group},
  author = {Schollw{\"o}ck, Ulrich},
  year = 2011,
  month = jul,
  journal = {Philosophical Transactions of the Royal Society A: Mathematical, Physical and Engineering Sciences},
  volume = {369},
  number = {1946},
  pages = {2643--2661},
  issn = {1364-503X, 1471-2962},
  doi = {10.1098/rsta.2010.0382},
  urldate = {2025-06-12},
  langid = {english}
}

@article{SCHOLLWOCK2011densitymatrixa,
  title = {The Density-Matrix Renormalization Group in the Age of Matrix Product States},
  author = {Schollw{\"o}ck, Ulrich},
  year = 2011,
  month = jan,
  journal = {Annals of Physics},
  volume = {326},
  number = {1},
  pages = {96--192},
  issn = {00034916},
  doi = {10.1016/j.aop.2010.09.012},
  urldate = {2025-06-12},
  copyright = {https://www.elsevier.com/tdm/userlicense/1.0/},
  langid = {english}
}

@misc{SHA2025Exponential,
  title = {Exponential Convergence of the Local Diabatic Representation for Nonadiabatic Models},
  author = {Sha, Mo and Gu, Bing},
  year = 2025,
  month = sep,
  number = {arXiv:2509.05694},
  eprint = {2509.05694},
  primaryclass = {physics},
  publisher = {arXiv},
  doi = {10.48550/arXiv.2509.05694},
  urldate = {2025-09-26},
  archiveprefix = {arXiv},
  langid = {english}
}

@article{SUBOTNIK2008Constructing,
  title = {Constructing Diabatic States from Adiabatic States: {{Extending}} Generalized {{Mulliken}}--{{Hush}} to Multiple Charge Centers with {{Boys}} Localization},
  shorttitle = {Constructing Diabatic States from Adiabatic States},
  author = {Subotnik, Joseph E. and Yeganeh, Sina and Cave, Robert J. and Ratner, Mark A.},
  year = 2008,
  journal = {Journal of Chemical Physics},
  volume = {129},
  number = {24},
  pages = {244101},
  publisher = {American Institute of Physics}
}

@article{SUZUKI1976Generalized,
  title = {Generalized Trotter's Formula and Systematic Approximants of Exponential Operators and Inner Derivations with Applications to Many-Body Problems},
  author = {Suzuki, Masuo},
  year = 1976,
  month = jun,
  journal = {Communications in Mathematical Physics},
  volume = {51},
  number = {2},
  pages = {183--190},
  issn = {0010-3616, 1432-0916},
  doi = {10.1007/BF01609348},
  urldate = {2025-10-24},
  copyright = {http://www.springer.com/tdm},
  langid = {english}
}

@book{TANNOR2007Introduction,
  title = {Introduction to {{Quantum Mechanics}}: {{A Time-Dependent Perspective}}},
  author = {Tannor, David J.},
  year = 2007,
  month = jan,
  publisher = {University Science Books},
  googlebooks = {t7m08j3Wi9YC},
  isbn = {978-1-891389-23-8}
}

@article{WANG2003Multilayer,
  title = {Multilayer Formulation of the Multiconfiguration Time-Dependent Hartree Theory},
  author = {Wang, Haobin and Thoss, Michael},
  year = 2003,
  month = jul,
  journal = {Journal of Chemical Physics},
  volume = {119},
  number = {3},
  pages = {1289--1299},
  issn = {0021-9606},
  doi = {10.1063/1.1580111},
  urldate = {2025-09-28},
  langid = {english}
}

@article{WANG2024Impact,
  title = {Impact of {{Geometric Phase}} on {{Dynamics}} of {{Complex-Forming Reactions}}: {{H}} + {{O2}} {$\rightarrow$} {{OH}} + {{O}}},
  shorttitle = {Impact of {{Geometric Phase}} on {{Dynamics}} of {{Complex-Forming Reactions}}},
  author = {Wang, Junyan and Xie, Changjian and Hu, Xixi and Guo, Hua and Xie, Daiqian},
  year = 2024,
  month = apr,
  journal = {The Journal of Physical Chemistry Letters},
  volume = {15},
  number = {16},
  pages = {4237--4243},
  publisher = {American Chemical Society},
  doi = {10.1021/acs.jpclett.4c00789},
  urldate = {2025-08-22}
}

@article{WORNER2011Conical,
  title = {Conical {{Intersection Dynamics}} in {{NO2 Probed}} by {{Homodyne High-Harmonic Spectroscopy}}},
  author = {W{\"o}rner, H. J. and Bertrand, J. B. and Fabre, B. and Higuet, J. and Ruf, H. and Dubrouil, A. and Patchkovskii, S. and Spanner, M. and Mairesse, Y. and Blanchet, V. and M{\'e}vel, E. and Constant, E. and Corkum, P. B. and Villeneuve, D. M.},
  year = 2011,
  month = oct,
  journal = {Science},
  volume = {334},
  number = {6053},
  pages = {208--212},
  publisher = {American Association for the Advancement of Science},
  doi = {10.1126/science.1208664},
  urldate = {2022-05-12}
}

@inbook{WORTH2004Multidimensional,
  title = {Multidimensional Dynamics Involving a Conical Intersection: Wavepacket Calculations Using the Mctdh Method},
  shorttitle = {Multidimensional Dynamics Involving a Conical Intersection},
  booktitle = {Advanced {{Series}} in {{Physical Chemistry}}},
  author = {Worth, G. A. and Meyer, H.-D. and Cederbaum, L. S.},
  year = 2004,
  month = jul,
  volume = {15},
  pages = {583--617},
  publisher = {WORLD SCIENTIFIC},
  doi = {10.1142/9789812565464_0014},
  urldate = {2025-09-28},
  collaborator = {Domcke, Wolfgang and Yarkony, David R and K{\"o}ppel, Horst},
  isbn = {978-981-238-672-4 978-981-256-546-4},
  langid = {english}
}

@article{XIE2016Nonadiabatic,
  title = {Nonadiabatic {{Tunneling}} in {{Photodissociation}} of {{Phenol}}},
  author = {Xie, Changjian and Ma, Jianyi and Zhu, Xiaolei and Yarkony, David R. and Xie, Daiqian and Guo, Hua},
  year = 2016,
  month = jun,
  journal = {Journal of the American Chemical Society},
  volume = {138},
  number = {25},
  pages = {7828--7831},
  publisher = {American Chemical Society},
  issn = {0002-7863},
  doi = {10.1021/jacs.6b03288},
  urldate = {2025-08-20}
}

@article{XIE2017Constructive,
  title = {Constructive and {{Destructive Interference}} in {{Nonadiabatic Tunneling}} via {{Conical Intersections}}},
  author = {Xie, Changjian and Kendrick, Brian K. and Yarkony, David R. and Guo, Hua},
  year = 2017,
  month = may,
  journal = {Journal of Chemical Theory and Computation},
  volume = {13},
  number = {5},
  pages = {1902--1910},
  publisher = {American Chemical Society},
  issn = {1549-9618},
  doi = {10.1021/acs.jctc.7b00124},
  urldate = {2025-08-20}
}

@article{XIE2024NondirectProduct,
  title = {Nondirect-{{Product Local Diabatic Representation}} with {{Smolyak Sparse Grids}}},
  author = {Xie, Yujuan and Yang, Yukun and Zhu, Xiaotong and Chen, Ahai and Gu, Bing},
  year = 2024,
  month = nov,
  journal = {Journal of Chemical Theory and Computation},
  volume = {20},
  number = {21},
  pages = {9512--9521},
  publisher = {American Chemical Society},
  issn = {1549-9618},
  doi = {10.1021/acs.jctc.4c00673},
  urldate = {2025-06-12},
  langid = {english}
}

@article{XIE2025Linked,
  title = {Linked Product Approximation to the Global Electronic Overlap Matrix},
  author = {Xie, Yujuan and Gu, Bing},
  year = 2025,
  month = sep,
  journal = {Journal of Chemical Theory and Computation},
  volume = {21},
  number = {19},
  pages = {9249--9258},
  publisher = {American Chemical Society},
  issn = {1549-9618},
  doi = {10.1021/acs.jctc.5c00862},
  urldate = {2025-09-29},
  langid = {english}
}

@article{XIE2025Quantum,
  title = {Quantum Geometrical Molecular Dynamics},
  author = {Xie, Yujuan and Liu, Ruoxi and Gu, Bing},
  year = 2025,
  month = dec,
  journal = {Science Advances},
  volume = {11},
  number = {50},
  pages = {eadz3711},
  publisher = {American Association for the Advancement of Science},
  doi = {10.1126/sciadv.adz3711},
  urldate = {2026-01-08}
}

@article{YARKONY2019Diabatic,
  title = {Diabatic and Adiabatic Representations: {{Electronic}} Structure Caveats},
  shorttitle = {Diabatic and Adiabatic Representations},
  author = {Yarkony, David R. and Xie, Changjian and Zhu, Xiaolei and Wang, Yuchen and Malbon, Christopher L. and Guo, Hua},
  year = 2019,
  month = mar,
  journal = {Computational and Theoretical Chemistry},
  volume = {1152},
  pages = {41--52},
  issn = {2210-271X},
  doi = {10.1016/j.comptc.2019.01.020},
  urldate = {2021-07-27},
  langid = {english}
}

@article{YUAN2018Observation,
  title = {Observation of the Geometric Phase Effect in the {{H}} + {{HD}} {$\rightarrow$} {{H2}} + {{D}} Reaction},
  author = {Yuan, Daofu and Guan, Yafu and Chen, Wentao and Zhao, Hailin and Yu, Shengrui and Luo, Chang and Tan, Yuxin and Xie, Ting and Wang, Xingan and Sun, Zhigang and Zhang, Dong H. and Yang, Xueming},
  year = 2018,
  month = dec,
  journal = {Science},
  volume = {362},
  number = {6420},
  pages = {1289--1293},
  publisher = {American Association for the Advancement of Science},
  doi = {10.1126/science.aav1356},
  urldate = {2023-03-30}
}

@misc{ZHANG2020stability,
  title = {On Stability of Tensor Networks and Canonical Forms},
  author = {Zhang, Yifan and Solomonik, Edgar},
  year = 2020,
  month = jan,
  number = {arXiv:2001.01191},
  eprint = {2001.01191},
  primaryclass = {math},
  publisher = {arXiv},
  doi = {10.48550/arXiv.2001.01191},
  urldate = {2025-10-24},
  archiveprefix = {arXiv},
  langid = {english}
}

@article{ZHOU2019QuasiDiabatic,
  title = {Quasi-{{Diabatic Scheme}} for {{Nonadiabatic On-the-Fly Simulations}}},
  author = {Zhou, Wanghuai and Mandal, Arkajit and Huo, Pengfei},
  year = 2019,
  month = nov,
  journal = {Journal of Physical Chemistry Letters},
  volume = {10},
  number = {22},
  pages = {7062--7070},
  publisher = {American Chemical Society},
  doi = {10.1021/acs.jpclett.9b02747},
  urldate = {2023-06-29}
}

@article{ZHU2015Construction,
  title = {On the {{Construction}} of {{Property Based Diabatizations}}: {{Diabolical Singular Points}}},
  shorttitle = {On the {{Construction}} of {{Property Based Diabatizations}}},
  author = {Zhu, Xiaolei and Yarkony, David R.},
  year = 2015,
  month = dec,
  journal = {Journal of Physical Chemistry A},
  volume = {119},
  number = {50},
  pages = {12383--12391},
  publisher = {American Chemical Society},
  issn = {1089-5639},
  doi = {10.1021/acs.jpca.5b07705},
  urldate = {2024-03-11}
}

@article{ZHU2024Making,
  title = {Making {{Peace}} with {{Random Phases}}: {{Ab Initio Conical Intersection Quantum Dynamics}} in {{Random Gauges}}},
  shorttitle = {Making {{Peace}} with {{Random Phases}}},
  author = {Zhu, Xiaotong and Gu, Bing},
  year = 2024,
  month = aug,
  journal = {The Journal of Physical Chemistry Letters},
  volume = {15},
  number = {33},
  pages = {8487--8493},
  publisher = {American Chemical Society},
  issn = {1948-7185, 1948-7185},
  doi = {10.1021/acs.jpclett.4c01688},
  urldate = {2025-06-12},
  copyright = {https://doi.org/10.15223/policy-029},
  langid = {english}
}

\end{document}


\title{Supplementary Information for \\ ``Coarse-Grained Geometric Quantum Dynamics in the Tensor Network Representation''}
\maketitle

\section{Tensor Train Algorithm used in this work}\label{ttin}
\begin{algorithm}[H]
    \caption{Tensor-Train SVD (TT-SVD)}
    \label{alg:ttsvd}
    \begin{algorithmic}[1]
        \Require A $d$-dimensional tensor $\mathbf{A}$ of size $n_0 \times \cdots \times n_{d-1}$.
        \Statex \hspace{\algorithmicindent} Parameters: Decomposition error $\epsilon$, max bond dimension $D_{\max}$.
        \Ensure TT-cores $\mathbf{A}^{[0]}, \dots, \mathbf{A}^{[d-1]}$.
        
        \State $\mathbf{C}_{\text{res}} \leftarrow \mathbf{A}$ \Comment{Initialize residual tensor}
        \State $D_{\text{left}} \leftarrow 1$      
        \For{$k = 0$ to $d-2$}
            \State $\mathbf{M} \leftarrow \text{reshape}(\mathbf{C}_{\text{res}}, [D_{\text{left}} \cdot n_k, -1])$  
            
            \State $(\mathbf{U}, \boldsymbol{\Sigma}, \mathbf{V}) \leftarrow \text{TruncatedSVD}(\mathbf{M}, \delta=\epsilon/\sqrt{d-1}, D_{\max})$ \Comment{Detailed in \cref{alg:trunc_svd}}
            
            \State $D_{\text{right}} \leftarrow \text{size}(\boldsymbol{\Sigma}, 1)$
            \State $\mathbf{A}^{[k]} \leftarrow \text{reshape}(\mathbf{U}, [D_{\text{left}}, n_k, D_{\text{right}}])$
            \State $\mathbf{C}_{\text{res}} \leftarrow \boldsymbol{\Sigma} \mathbf{V}^T$
            \State $D_{\text{left}} \leftarrow D_{\text{right}}$
        \EndFor
        \State $\mathbf{A}^{[d-1]} \leftarrow \mathbf{C}_{\text{res}}$
    \end{algorithmic}
\end{algorithm}

\begin{algorithm}[H]
    \caption{Truncated SVD Procedure}
    \label{alg:trunc_svd}
    \begin{algorithmic}[1]
        \Require Matrix $\mathbf{M}$.
        \Statex \hspace{\algorithmicindent} Parameters:Local truncation error $\delta$, max rank $D_{\max}$.
        \Ensure Truncated error $\mathbf{U}', \boldsymbol{\Sigma}', \mathbf{V}'$.
        
        \State Perform SVD: $\mathbf{M} = \mathbf{U} \boldsymbol{\Sigma} \mathbf{V}^T$ with singular values $\sigma_0 \ge \sigma_1 \ge \dots \ge 0$.
        
        \State Find the smallest integer $D$ such that $\sqrt{\sum_{j=D}^{\text{end}} \sigma_j^2} \le \delta$.\label{step:svd_trunc_start}
        
        \If{$D_{\max}$ is provided}
            \State $D \leftarrow \min(D, D_{\max})$ 
        \EndIf\label{step:svd_trunc_end}
        
        \State $\mathbf{U}' \leftarrow \mathbf{U}(:, 0:D)$
        \State $\boldsymbol{\Sigma}' \leftarrow \boldsymbol{\Sigma}(0:D, 0:D)$
        \State $\mathbf{V}' \leftarrow \mathbf{V}(:, 0:D)$
        
        \State \Return $\mathbf{U}', \boldsymbol{\Sigma}', \mathbf{V}'$
    \end{algorithmic}
\end{algorithm}

\begin{algorithm}[H]
    \caption{TT-rounding}
    \label{alg:ttround}
    \begin{algorithmic}[1]
        \Require A $d$-dimensional left-canonical tensor train $\mathbf{A}$.
        \Statex \hspace{\algorithmicindent} Parameters: Truncation error $\epsilon$, max bond dimension $D_{\max}$.
        \Ensure Compressed tensor train $\mathbf{A'}$.

        \State $\mathbf{A'} \leftarrow \mathbf{A}$
        \For {$k=d-1$ down to $1$}
            \State $D_k, n_k, D_{k+1} \leftarrow \text{size}(\mathbf{A'}^{[k]})$
            \State $\mathbf{M} \leftarrow \text{reshape}(\mathbf{A'}^{[k]}, [D_k, n_k \cdot D_{k+1}])$
            
            \State $(\mathbf{U}, \boldsymbol{\Sigma}, \mathbf{V}) \leftarrow \text{TruncatedSVD}(\mathbf{M}, \delta=\epsilon/\sqrt{d-1}, D_{\max})$ \Comment{Detailed in \cref{alg:trunc_svd}}
            
            \State $D'_{k} \leftarrow \text{size}(\boldsymbol{\Sigma}, 1)$
            \State $\mathbf{A'}^{[k]} \leftarrow \text{reshape}(\mathbf{V}^T, [D'_{k}, n_k, D_{k+1}])$
            
            \State $\mathbf{A'}^{[k-1]} \leftarrow \text{reshape}(\mathbf{A'}^{[k-1]}, [-1, D_k]) \cdot (\mathbf{U} \boldsymbol{\Sigma})$ \Comment{Absorb U into previous core}
            \State $\mathbf{A'}^{[k-1]} \leftarrow \text{reshape}(\mathbf{A'}^{[k-1]}, [D_{k-1}, n_{k-1}, D'_{k}])$
        \EndFor
        
        \State \Return $\mathbf{A'}$
    \end{algorithmic}
\end{algorithm}

The accuracy of TT-SVD and TT-rounding is governed by the singular value truncation at each step. The local truncation error is quantified by the square root of the sum of the squares of the discarded singular values~\cite{ECKART1936Approximation, MIRSKY1960SYMMETRIC, OSELEDETS2011TensorTrain}:
\begin{equation}\label{eq:svd_abs_error}
    \delta = \sqrt{\sum_{j=D} \sigma_j^2},
\end{equation}
where $\sigma_j$ are the singular values arranged in descending order. A uniform local error threshold $\delta$ yields a total error of $\epsilon = \delta\sqrt{d-1}$. However, slowly decaying singular values—typical in strongly entangled systems—can lead to excessive bond dimensions. To ensure computational feasibility, we employ a hybrid truncation strategy as detailed in Algorithm~\ref{alg:trunc_svd}. The retained rank $D$ is first determined by the error threshold $\delta$ and subsequently capped by a hard limit $D_{\max}$, such that the final rank is given by $\min(D, D_{\max})$.

\newpage

\section{Construction of the propagators in the tensor train representation}\label{buildpropagator}

\begin{figure}[htbp]
    \includegraphics[width=1\columnwidth]{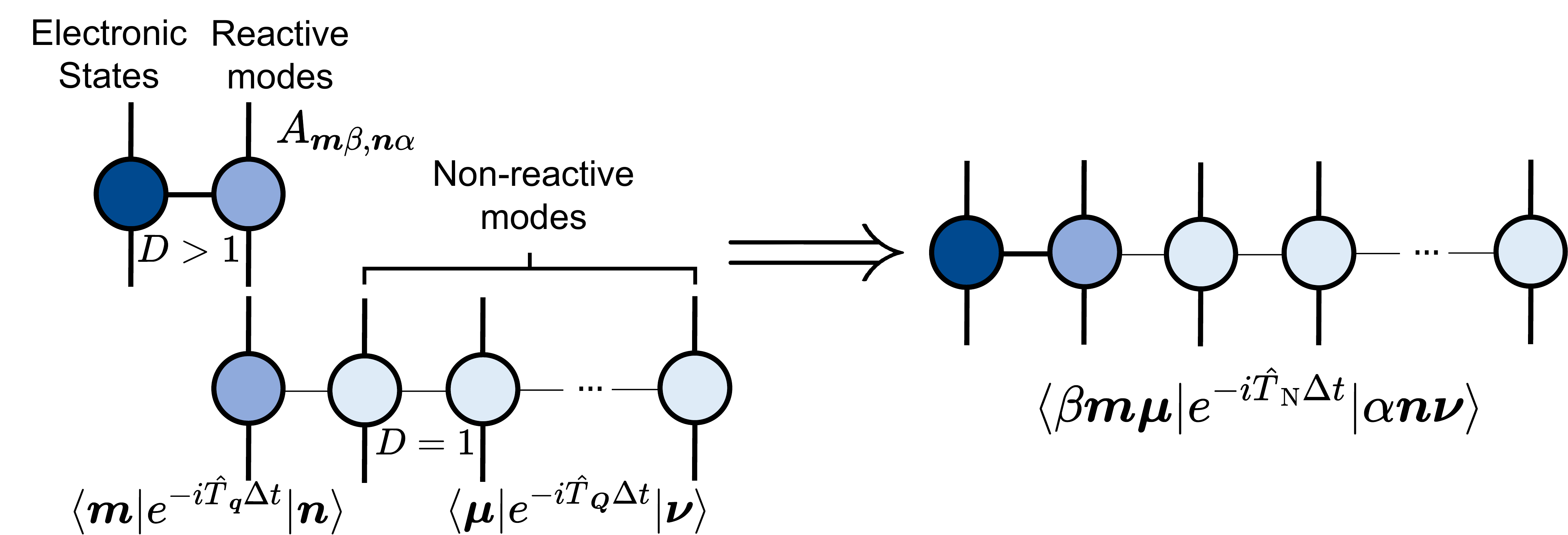}
    \caption{Schematic construction of the kinetic energy propagator $e^{-i\hat{T}_\text{N}\Delta t}$ in the coarse-grained method. The line width schematically represents the corresponding bond dimension. Different colors represent different types of indices.}
    \label{fig:kinetic}
\end{figure}

\begin{figure}[htbp]
    \includegraphics[width=1\columnwidth]{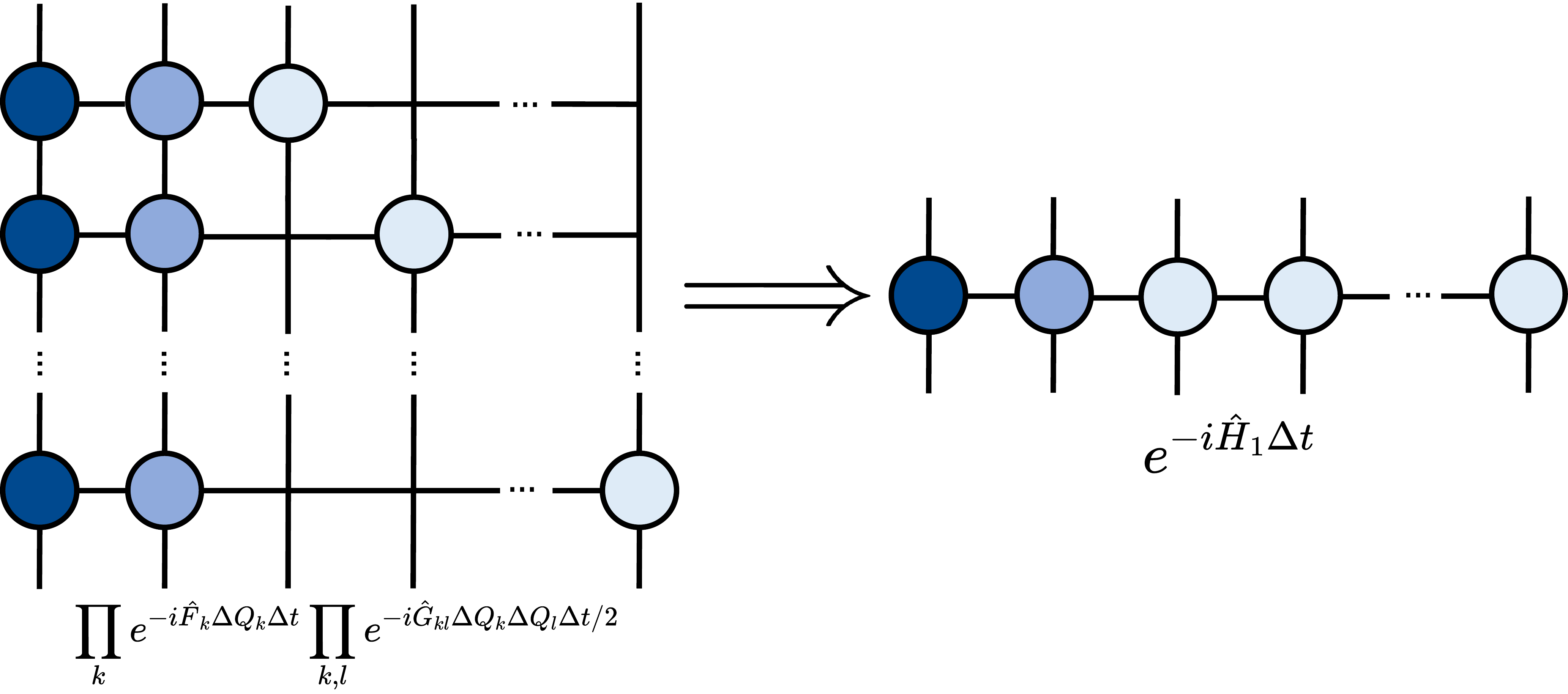}
    \caption{Schematic construction of the propagator $e^{-i\hat{H}_{1}\Delta t}$ in the coarse-grained method. For clarity, only $\prod_{k} e^{-i\hat{F}_k \Delta Q_{k} {\Delta t}}$ is shown. The line width schematically represents the corresponding bond dimension. The empty intersections correspond to identity matrices at the corresponding sites. Different colors represent different types of indices.}
    \label{fig:H1}
\end{figure}

\begin{figure}[htbp]
    \includegraphics[width=0.6\columnwidth]{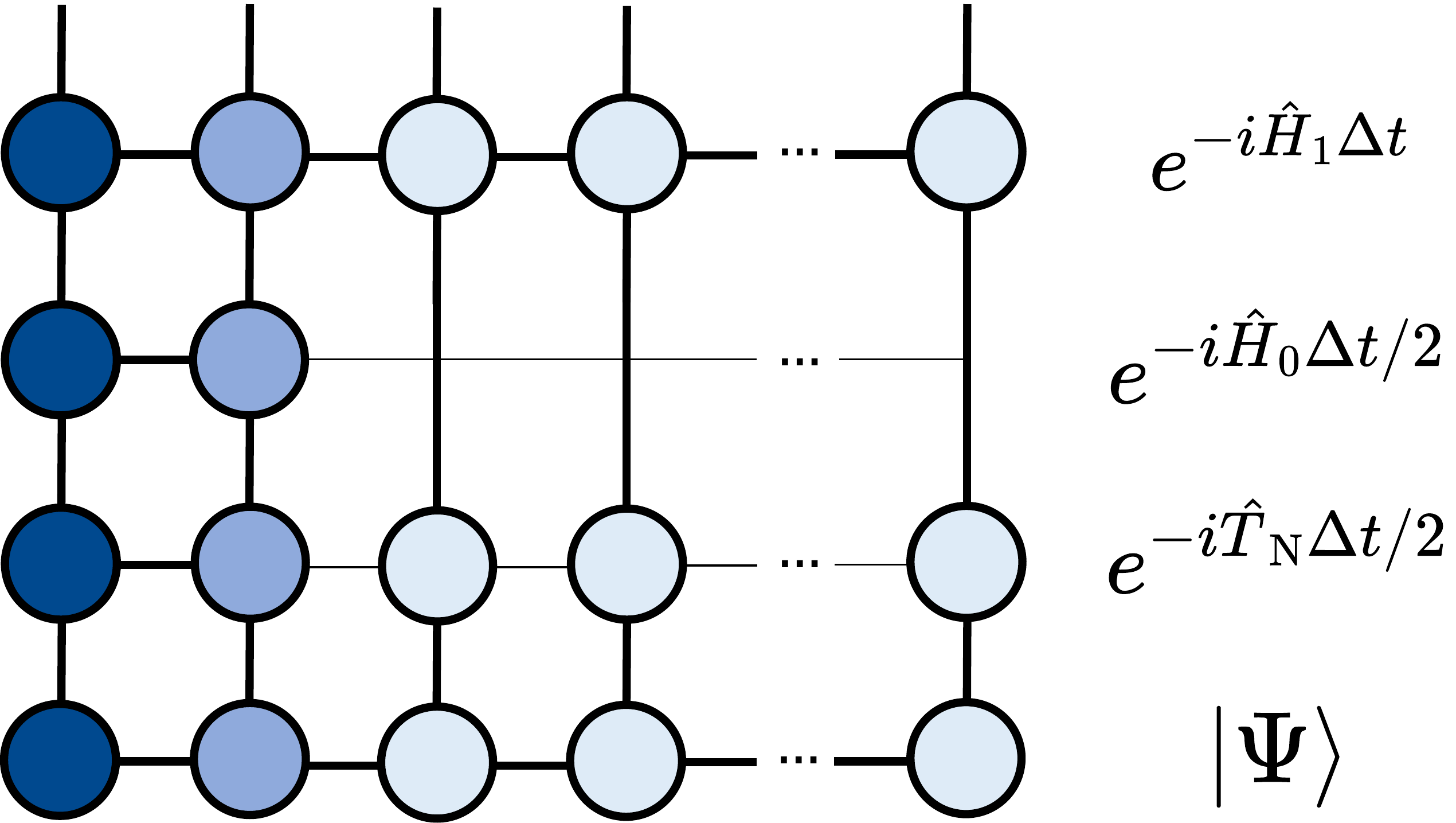}
    \caption{Schematic construction of a half-step evolution of the wavefunction in the coarse-grained method. The line width schematically represents the corresponding bond dimension. $|\psi\rangle$ represents the wavefunction at the current time step. Different colors represent different types of indices.}
    \label{fig:evolution}
\end{figure}

\newpage

\section{Vibronic coupling model for pyrazine}
The vibronic coupling Hamiltonian is given by
\begin{equation}\label{modelH}
	\begin{aligned}
		\hat{H}=\sum_i{\frac{\omega_i}{2}\qty(-\frac{\partial^2}{\partial {R_i}^2}+\hat{R_i}^2)}\mathbf{I}+
		\begin{bmatrix}
			-\Delta&0\\
			0&\Delta
		\end{bmatrix}
		+\sum_{i\in G_1} \begin{bmatrix}
			a_i&0\\
			0&b_i
		\end{bmatrix}\hat{R_i}
		+\sum_{i\in G_3} \begin{bmatrix}
			0&c_i \\
			c_i&0
		\end{bmatrix}\hat{R_i}\\
		+\sum_{(i,j)\in G_2} \begin{bmatrix}    
			a_{i,j}&0\\
			0&b_{i,j}
		\end{bmatrix}\hat{R_i} \hat{R_j}
		+\sum_{(i,j)\in G_4} \begin{bmatrix}
			0&c_{i,j}\\
			c_{i,j}&0
		\end{bmatrix}\hat{R_i} \hat{R_j}
	\end{aligned}
\end{equation}
where $\mathbf{I}$ is the identity matrix, $\omega_i$ denotes the ground-state
vibrational frequency of the $i$-th normal mode, $R_i$ is the
corresponding dimensionless normal-mode coordinate. $G_i$ refers to four different symmetry groups of the vibrational modes.
The parameters are extracted from reference\cite{RAAB1999Molecular}.

\section{Error Analysis}
The error in the coarse-grained method arises from three main sources: the Trotter error from the time-stepping algorithm ($\mathcal{O}(\Delta t^3)$), the truncation error in TT-SVD and TT-rounding, and the error originating from the coarse-grained ansatz.
The Trotter error can be systematically controlled by reducing the time step $\Delta t$. In this section, we focus on analyzing the truncation error and the coarse-graining error.

\begin{figure}[h!]
    \centering
    \includegraphics[width=1\textwidth]{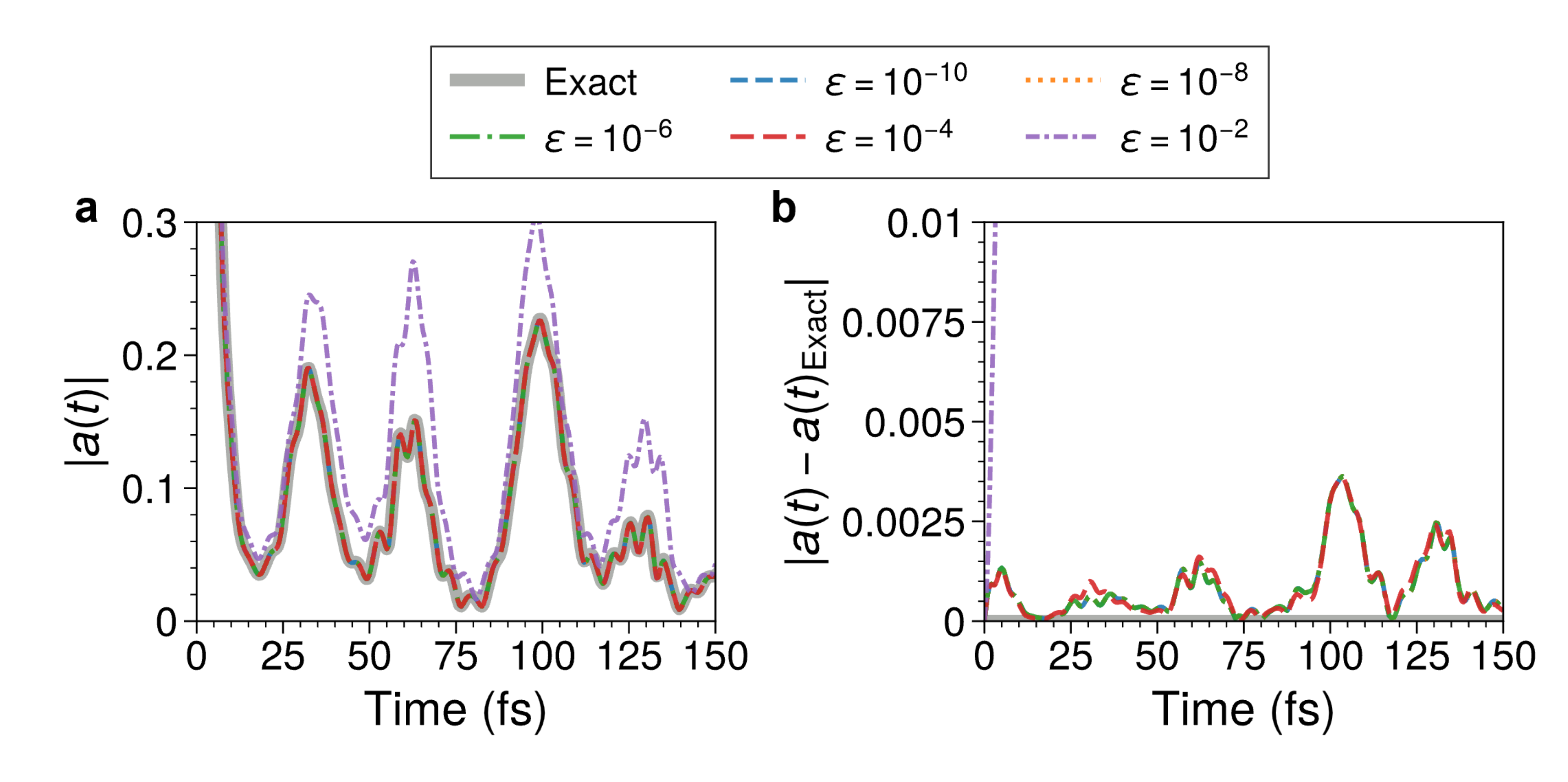}
    \caption{Error analysis of the calculation with respect to the truncation error $\epsilon$. (a) The absolute value of the autocorrelation function, $|a(t)|$, calculated with different $\epsilon$ values, compared to the exact result. (b) The absolute difference in $|a(t)|$ between the TT calculations and the exact result.}
    \label{Fig:epsilon_test}
\end{figure}

First, we investigate the effect of the truncation error parameter $\epsilon$. We set the maximum bond dimension $D_\text{max}=200$, ensuring that the truncation error is predominantly governed by $\epsilon$.
\cref{Fig:epsilon_test} demonstrates that the autocorrelation function, defined as $a(t) = \langle \Psi(0)|\Psi(t)\rangle$, converges rapidly with respect to this parameter. A threshold of $\epsilon=10^{-6}$ is sufficient to achieve excellent agreement. The simulation with this parameter required approximately 137 seconds of wall time.
The maximum RAM required, assuming the ``complex128'' data type, was approximately 40~GB, while the compressed file containing states and operators occupied less than 200~MB. The bond dimension reached a maximum of 98 at the virtual bond connecting the reactive and non-reactive coordinates.

\begin{figure}[h!]
    \includegraphics[width=1\textwidth]{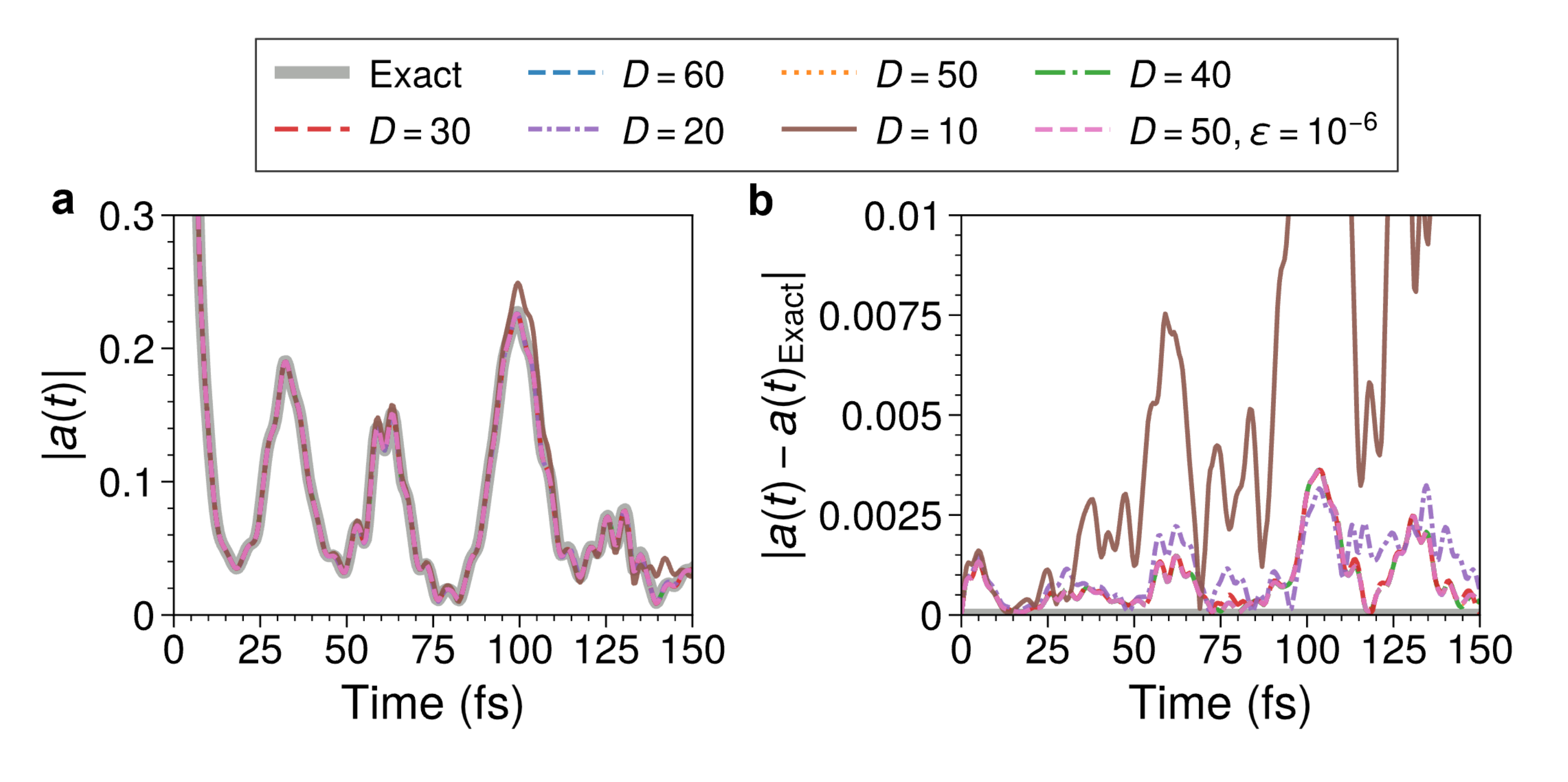}
    \caption{Error analysis of the calculation with respect to different maximum bond dimension $D_{\max}$. (a) The absolute value of the autocorrelation function, $|a(t)|$, calculated with different $D_{\max}$ values, compared to the exact result. (b) The absolute difference in $|a(t)|$ between the TT calculations and the exact result.}
    \label{Fig:Dmax_test}
\end{figure}

Next, we test the convergence with respect to $D_{\max}$. To isolate its effect, we set the error threshold $\epsilon$ to a value near machine precision ($10^{-14}$), ensuring that truncation is exclusively determined by $D_{\max}$. As shown in \cref{Fig:Dmax_test}, increasing $D_{\max}$ systematically improves accuracy, with the autocorrelation function fully converging for $D_{\max}$ in the range of 40 to 50. For the case of $D_{\max}=40$, the simulation required approximately 8000 seconds of wall time, with a theoretical peak transient memory of approximately 17~GB for the complex128 wavefunction, while the size of the saved wavepacket file at each time step was approximately 120~MB. However, by introducing a practical error threshold of $\epsilon=10^{-6}$ with a sufficient $D_{\max}=50$, a result of comparable accuracy is obtained (see \cref{Fig:Dmax_test}), while the computational cost is dramatically reduced: the approximate wall time decreases to 152 seconds, with a peak transient memory of approximately 3~GB and a file size per time step of approximately 143~MB. Notably, the rapid convergence with respect to $D_{\max}$ suggests that the 4-mode pyrazine model does not generate high levels of entanglement, making it particularly well-suited for the TT approach.

\begin{figure}[h!]
    \centering
    \includegraphics[width=1\textwidth]{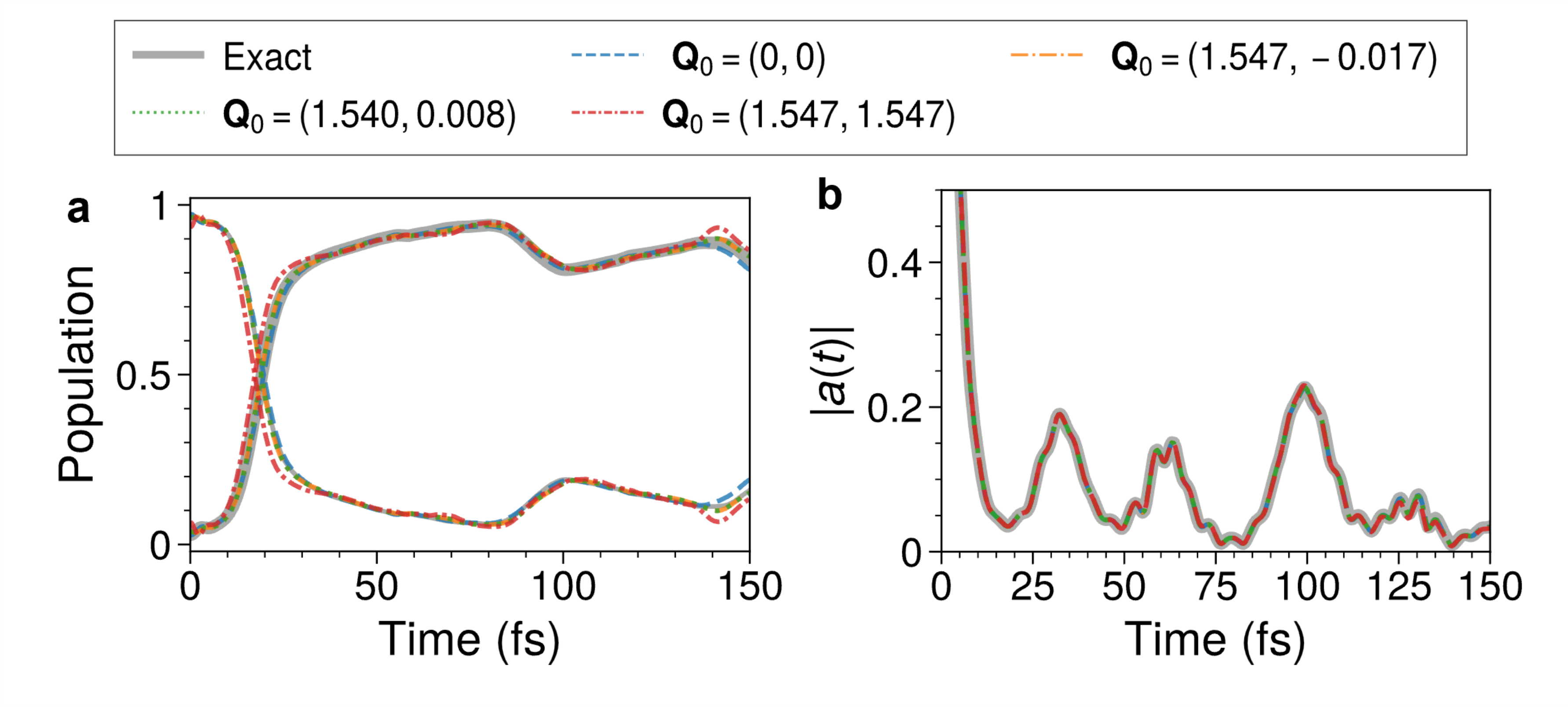}
    \caption{Influence of the reference geometry $\mathbf{Q}_0$ on the TT calculation. (a) Time evolution of the electronic population of the $S_2$ state for four different choices of $\mathbf{Q}_0$, compared to the exact result. (b) The absolute difference in the autocorrelation function $|a(t)|$ relative to the exact result for each choice of $\mathbf{Q}_0$.}
    \label{Fig:Q0_test}
\end{figure}

Finally, we assess the influence of the reference geometry for non-reactive coordinates, $\mathbf{Q}_0$, a critical parameter in our coarse-graining approach. As the reference point for the Taylor expansion of the Hamiltonian, its choice directly determines the quality of the approximated potential energy surfaces. To investigate this dependence, we performed simulations using four distinct geometries:
\begin{enumerate}[label=(\roman*)]
    \item The ground state minimum point $\mathbf{Q}_0 = (0,0)$.
    \item The projection of the minimum energy point of the conical intersection hypersurface for the 4-mode model, $\mathbf{Q}_0 = (1.547, -0.017)$, for modes $(\nu_1, \nu_{9a})$~\cite{RAAB1999Molecular}.
    \item The projection of the minimum energy point of the conical intersection hypersurface for the 24-mode model, $\mathbf{Q}_0 = (1.540, 0.008)$, for modes $( \nu_1, \nu_{9a})$~\cite{RAAB1999Molecular}.
    \item An arbitrarily chosen geometry $\mathbf{Q}_0=(1.547,1.547)$.
\end{enumerate}
The choice of the reference point $\mathbf{Q}_0$ influences the overall relaxation dynamics, with different physically motivated geometries performing better for different observables. Setting $\mathbf{Q}_0$ at the Franck-Condon point (i) yields a more accurate autocorrelation function, whereas choosing the conical intersection minima (ii and iii) results in more accurate population dynamics. In contrast, the arbitrary geometry (iv) produces large discrepancies in both autocorrelation and population dynamics. Hence, the predictive accuracy of the method depends sensitively on selecting a physically appropriate reference geometry.

\newpage
\section{Long-time Dynamics}

\begin{figure}[htbp]
    \centering
    \includegraphics[width=0.6\textwidth]{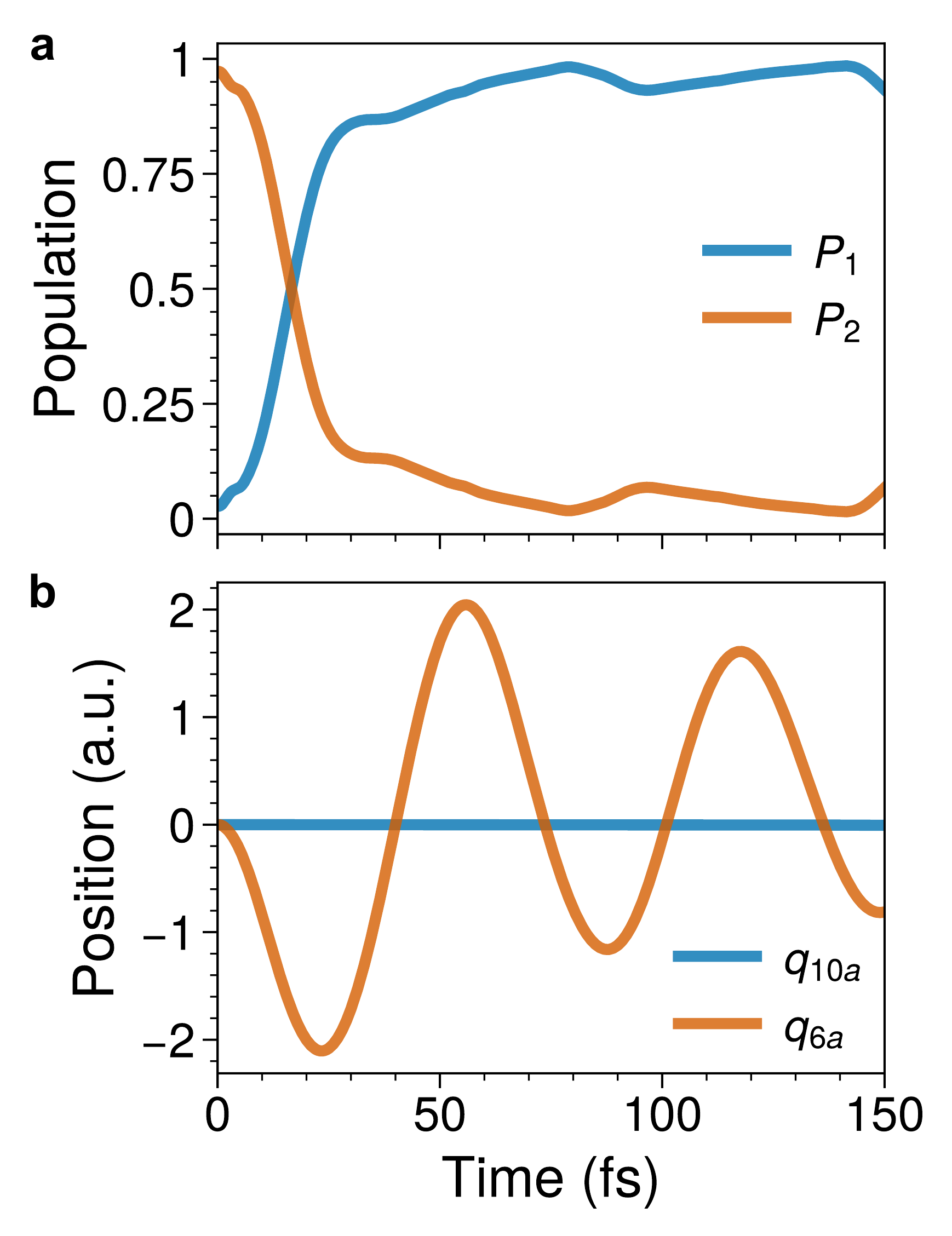}
    \caption{Non-adiabatic dynamics of the full 24-mode pyrazine model over \fs{150}. {(a)} Electronic population dynamics of the initially excited state $S_2$ and the lower-lying state $S_1$. {(b)} Position expectation values for the coupling mode $q_{10a}$ and the tuning mode $q_{6a}$.}
    \label{Fig:full24modelonger}
\end{figure}

\appendix

\bibliography{all}